\documentclass[12pt,letterpaper]{JHEP3}

\usepackage{amsmath,amssymb}
\usepackage{array}
\usepackage{slashed}

\usepackage{float}
\usepackage{amsmath,amsfonts,graphicx,bm}

\usepackage{amsmath,amsfonts,graphicx,bbm,multicol}

\usepackage{subfigure}

\newcommand\rep\mathbf

\def\beq{\begin{equation}}
\def\eeq{\end{equation}}
\def\bea{\begin{eqnarray}}
\def\eea{\end{eqnarray}}

\DeclareMathOperator\Tr{Tr}

\preprint{MADPH--10-1563}

\title{\vspace*{.75in}
Colored Resonant Signals at the LHC: \\
 Largest Rate and Simplest Topology }

\author{Tao Han,  Ian Lewis, and
Zhen Liu\thanks{ than@hep.wisc.edu,\ ilewis@wisc.edu,\ zliu57@wisc.edu}\\
{\it Department of Physics, 1150 University Avenue,
University of Wisconsin, Madison, WI 53706, U.S.A.} }

\abstract{
We study the colored resonance production at the LHC in a most general approach.
We classify the possible colored resonances based on group theory decomposition,
and construct their effective interactions with light partons.
The production cross section from annihilation of valence quarks or gluons
may be on the order of  $400-1000$ pb at LHC energies for a mass of 1 TeV
with nominal couplings,
leading to the largest production rates for new physics at the TeV scale,
and simplest event topology with dijet final states.
We apply the new dijet data from the LHC experiments to put bounds on various possible
colored resonant states. The current bounds range from $0.9$ to $2.7$ TeV.
The formulation is readily applicable for future searches including other decay modes.
}

\begin{document}

\section{Introduction}

With the first crop of data being released, the CERN Large Hadron Collider (LHC)
is already pushing the energy frontier and taking the field of high energy physics to a new era.
While much of the attention for new physics discovery has centered on theories
associated with electroweak symmetry breaking,
most initial states at hadron colliders are composed of colored particles.
Hence, any new colored resonances will be produced with favorable rates at the LHC
since their couplings may be typically of the strength of the strong-interaction.

Beyond the Standard Model (SM), there are many possible exotic colored states that can be produced
at the LHC.  Besides being phenomenologically interesting and experimentally
important to search for,  many of the exotic states are also theoretically motivated.
For example, color-antitriplet scalars may be produced via quark-quark annihilation
as squarks in R-parity violating supersymmetric (SUSY) theories \cite{Barbier:2004ez},
or as ``diquarks'' in $E_{6}$ grand unified theories \cite{Hewett:1988xc}.
Color-sextet scalars can arise in partially unified Pati-Salam theories \cite{Pati:1974yy} 
and be produced also via quark-quark annihilation.
Color-triplet fermions can be produced via quark-gluon annihilation as ``excited quarks" 
in composite models \cite{Cabibbo:1983bk,Baur:1987ga}.
Sextet fermions, the so-called ``quixes'',  associated 
with chiral color \cite{Frampton:1987dn} and top quark condensate models \cite{Martin:1992aq}
may also be produced via quark-gluon annihilation.
Color-octet scalars that are $SU(2)_L$ singlets can arise in technicolor models \cite{Hill:2002ap},
and in universal extra dimensions \cite{Dobrescu:2007xf}.
Color-octet vectors have been extensively explored as
axigluons \cite{Frampton:1987dn,Bagger:1987fz} and colorons \cite{Hill:1991at,Chivukula:1996yr}.
There has also been much recent interest in studying the similar states in the context of 
Kaluza-Klein gluons \cite{Agashe:2006hk}, and low-scale string 
resonances \cite{Cullen:2000ef,Burikham:2004su,Anchordoqui:2008hi} 
via  gluon-gluon, or quark-antiquark annihilation.

Any new resonant states produced at the LHC through interactions with light partons will  contribute
to the dijet production, leading to one of the simplest signal topologies.
Both the ATLAS and CMS
collaborations have recently searched for this class of signal and obtained $95\%$
confidence level limits on the production cross section of such resonant states.
From these limits they were able to already put the most stringent
bound on the mass of an excited quark of 1.53 TeV from ATLAS~\cite{Collaboration:2010bc} and 1.58 from CMS~\cite{Collaboration:2010jd},
and on a string resonance \cite{Collaboration:2010jd} of 2.5 TeV.

Motivated by the above considerations, we study the colored resonances in a most general way.
We classify them according to their couplings to light partons, solely
based on group theory decomposition as shown in section \ref{sec:class}.
Among those possible colored resonances, we focus on those produced by the leading
parton luminosities directly from valence quarks or gluons.
We then construct their couplings to light partons and describe their general features in section \ref{sec:inter}.
In section \ref{sec:prodrate} we calculate the cross sections for their resonant production
at the LHC with c.m.~energies of 7 and 14 TeV.
We apply the new ATLAS and CMS data to put bounds on various possible colored resonant states
in section \ref{sec:bnds}. Finally, we conclude in section \ref{sec:conc}.
A few appendices
contain the QCD color treatment, and a list of Feynman rules for the resonance couplings to
the initial state light partons.

\section{Classification of Resonant Particles in Hadronic Collisions}
\label{sec:class}

The resonance structures can be classified according to the spin ($J$) and the quantum numbers
under the SM gauge group $SU(3)_{C}\times SU(2)_{L}\times U(1)_{Y}$.
We adopt a notation of group structure
\begin{equation}
(SU_3,SU_2)_{Q_e}^J,
\label{eq:note}
\end{equation}
where $Q_e$ indicates the electric charge ($T_{3}+Y$).

In $pp$ collisions at the LHC, we consider the dominant partons participating in the heavy resonance
production to be the valence quarks and gluons. We express them with our notation as
\begin{eqnarray}
\begin{array}{clc}
Q~~~~~ &  (\mathbf{3},\mathbf{2})_{2/3,-1/3}^{1/2} ~~~~~ &  {\rm Left-handed\ doublet} \\
U~~~~~ &  (\mathbf{3},\mathbf{1})_{2/3}^{1/2} ~~~~~     &  {\rm Right-handed\ singlet} \\
D~~~~~ &  (\mathbf{3},\mathbf{1})_{-1/3}^{1/2} ~~~~~       &  {\rm Right-handed\ singlet} \\
A~~~~~ &  (\mathbf{8},\mathbf{1})_{0}^{1}  ~~~~~ &{\rm vector }.
\end{array}
\end{eqnarray}
We can thus classify the single particle production via the annihilation of any two partons above.
Table \ref{qnum.tab} lists the quantum numbers of possible resonances
in our notation from two initial partons.
Since the LHC is a ``QCD machine'', it is natural to start primarily based on the
 $SU(3)_{C}$ quantum numbers of the two initial states. We thus have partonic collisions of
quark-quark: $\rep3\otimes\rep3$;
quark-gluon: $\rep3\otimes\rep8$;
gluon-gluon: $\rep8\otimes\rep8$;
and quark-antiquark: $\rep3\otimes\bar{\rep3}$.
\begin{table}[tb]
\caption{The $SU(3)_{C}\times SU(2)_{L}\times U(1)_{Y}$ quantum numbers and spins ($J$)
of possible resonant states created by initial state quarks and gluons.
The electric charge $(Q_{e}=T_3+Y)$ and baryon number $(B)$ carried by the two initial
state partons  are also provided.}
\begin{center}
\begin{tabular}{|c|c|c|c|c|c|c|}  \hline
 initial state& $J$  & $SU(3)_{C}$ & $SU(2)_L$   & $U(1)_Y$  & $|Q_{e}|$ & $B$ \\ \hline
 $QQ$            &0& $\overline{\rep3}\oplus{\rep6}$& $\rep1\oplus\rep3$ &${1\over3}$ & $4\over3$,$2\over3$,$1\over3$ & $ 2\over3 $\\ \hline
 $QU$            &1& $\overline{\rep3}\oplus{\rep6}$& $\rep2$            &${5\over6}$ & $4\over3$,$1\over3$  & $ 2\over3 $\\ \hline
 $QD$            &1& $\overline{\rep3}\oplus{\rep6}$& $\rep2$             &$-{1\over6}$ & $2\over3$,$1\over3$ & $ 2\over3 $\\ \hline
 $UU$            &0& $\overline{\rep3}\oplus{\rep6}$ & $\rep1$            &${4\over3}$ & $4\over3$  & $ 2\over3 $   \\ \hline
 $DD$            &0& $\overline{\rep3}\oplus{\rep6}$ & $\rep1$            &$-{2\over3}$ & $2\over3$  & $ 2\over3 $   \\ \hline
 $UD$            &0& $\overline{\rep3}\oplus{\rep6}$ & $\rep1$            &${1\over3}$ & $1\over3$  & $ 2\over3 $  \\ \hline\hline
 $QA$            &$1\over2$, $3\over2$& $\rep3\oplus\bar{\rep6}\oplus\rep15$ &$\rep2$  & $1\over6$ & $2\over3$,$1\over3$ & $ 1\over3 $\\ \hline
 $UA$            &$1\over2$, $3\over2$& $\rep3\oplus\bar{\rep6}\oplus\rep15$ &$\rep1$  & $2\over3$ & $2\over3$   & $ 1\over3 $  \\ \hline
 $DA$            &$1\over2$, $3\over2$& $\rep3\oplus\bar{\rep6}\oplus\rep15$ &$\rep1$  & $1\over3$ & $1\over3$   & $ 1\over3 $   \\ \hline\hline
 $AA$            &0, 1, 2& $\rep1\oplus\rep8\oplus\rep8\oplus\rep10\oplus\bar{\rep10}\oplus\rep27$ & $\rep1$ &$0$ & $0$ & $0$\\ \hline \hline
  $Q\bar{Q}$      &1& $\rep1\oplus \rep8$          & $\rep1\oplus\rep3$ &$0$   & $1,0$   & $ 0 $ \\ \hline
 $Q\bar{U}$      &0& $\rep1\oplus\rep8$           & $\rep2$            &$-{1\over2}$& $1,0$   & $ 0 $  \\ \hline
 $Q\bar{D}$      &0& $\rep1\oplus\rep8$           & $\rep2$            &$1\over2$ & $1,0$   & $ 0 $  \\ \hline
 $U\bar{U},\ D\bar{D}$      &1& $\rep1\oplus\rep8$           & $\rep1$            &$0$   & $0$     & $ 0 $  \\ \hline
 $U\bar{D}$      &1& $\rep1\oplus\rep8$           & $\rep1$            &$1$   & $1$     & $ 0 $  \\ \hline
\end{tabular}
\end{center}
\label{qnum.tab}
\end{table}

Possible spins and the electric charges are also given in Table~\ref{qnum.tab}.
In principle, neutral  particles may be further classified by the discrete symmetries according
to their parity (P), charge conjugation (C), and  CP properties if these quantum numbers
are  conserved in their interactions. 
We will discuss them in the next section.
In the last column,  we add baryon numbers $(B)$ carried by the initial state partons.
Depending on the underlying theory for the new resonances, baryon number may or may not be
conserved in their interactions. 


\section{Parton-Resonance Interactions}
\label{sec:inter}

We now construct  the interaction Lagrangians for the resonances and partons
guided by the SM gauge symmetry.
We limit our consideration only to those colored states listed in Table~\ref{qnum.tab}.
We will not postulate their
interactions with other particles (leptons, electroweak bosons, or even new particles beyond the SM).
Although an incomplete description for a resonant particle as a full interacting theory, 
this minimal approach is sufficient for evaluating the production rate at the LHC. Assuming these
interactions dominate, then their decay to dijets would also be the leading channel.
Furthermore, we will 
not consider higher dimensional color representations beyond $\rep8$ again
due to the minimality considerations.  Should
there exist a color ``{\bf 15}-tet" fermion, a simple calculation of the QCD beta-function
would indicate the loss of the asymptotic freedom of QCD \cite{Chivukula:1990di}.

A similar approach to ours has been carried out to construct the 
potentially  large signals at the early run of the LHC with minimal model input  \cite{Bauer:2009cc,Barger:2006hm}.  There has also been previous work on classifying exotic particles at the LHC \cite{Ma:1998pi}.

%

\subsection{$\rep3\otimes\rep3$}
\label{33.SEC}
At the LHC the valence-valence initial states consist of two quarks, $uu$, $dd$, or $ud$.   Hence, the production cross section of a heavy particle that couples to two quarks will receive an enhancement from the parton luminosity of the initial state.
As listed on the top section of Table \ref{qnum.tab}, such states can be color-antitriplets or sextets.
They also carry an exotic baryon number of $2/3$ (if $B$ is conserved) and thus are often referred to
``diquarks''.\footnote{This should not be confused with a possible two-light-quark bound state as ``diquark''.
We are talking about a new state at a TeV mass scale with a quantum number similar to two quarks.}
According to their electroweak (EW) quantum numbers under $SU(2)_{L}\otimes U(1)_{Y}$,
there are 6 such states. We denote them by the notation in Eq.~(\ref{eq:note}) as
\begin{eqnarray}
&& \Phi \sim  ({\rep3}\oplus\bar{\rep6},\rep3)^0_{-4/3,2/3,-1/3},\quad
\Phi_q \sim ({\rep3}\oplus\bar{\rep6},\rep1)^0_q \ \ (q=-1/3,\ 2/3,\ -4/3),
\nonumber \\
&&
V_{U}^\mu \sim ({\rep3}\oplus\bar{\rep6},\rep2)^1_{-1/3,-4/3}\quad
V_{D}^\mu \sim ({\rep3}\oplus\bar{\rep6},\rep2)^1_{2/3,-1/3}.
\label{eq:33}
\end{eqnarray}
We construct the gauge invariant Lagrangian as follows
\begin{eqnarray}
\mathcal{L}_{qqD}
& \sim & K^j_{ab}\ \left[ y_{\alpha\beta}\ \overline{Q^C_{\alpha a}}i\sigma_2{\Phi^j}Q_{\beta b} +
\kappa_{\alpha\beta}\ {\Phi^{j}_{-1/3}}\overline{Q^C_{\alpha a}}i\sigma_2Q_{\beta b} \right.   \nonumber\\
&& + \lambda^{1/3}_{\alpha\beta}\ \Phi^{j}_{-1/3}\overline{D^C_{\alpha a}}U_{\beta b}
+\lambda^{2/3}_{\alpha\beta}\ \Phi^{j}_{2/3}\overline{D^{C}_{\alpha a}}D_{\beta b}
+\lambda^{4/3}_{\alpha\beta}\ \Phi^j_{-4/3}\overline{U^C_{\alpha a}}U_{\beta b} \nonumber\\
&&
\left. +\lambda^U_{\alpha\beta}\ \overline{Q^C_{\alpha a}}i\sigma_2\gamma_\mu{V^{j}_U}^\mu U_{\beta b}
+\lambda^D_{\alpha\beta}\ \overline{Q^{C}_{\alpha a}}i\sigma_2\gamma_\mu{V^{j}_D}^\mu D_{\beta b} \right] +\rm{h.c.},
\label{diquark.EQ}
\end{eqnarray}
where $\Phi^j = {1\over 2}\sigma_{k} \Phi_{k}^{j}$ with $\sigma_{k}$ the $SU(2)_{L}$ Pauli matrices and
$K^j_{ab}$ are $SU(3)_{C}$ Clebsch-Gordan coefficients with  the quark color indices
$a,b=1-3$, and the diquark color index $j=1-N_D$.   $N_D$ is the dimension of the ($N_D=3$) triplet or ($N_D=6$) antisextet representation.
$C$ denotes charge conjugation, and $\alpha,\beta$ are the fermion generation indices.
The color factor $K^j_{ab}$ is symmetric (antisymmetric) under $ab$ for the ${\rep{6}}\ (\bar{\rep{3}})$ representation.
Their normalization convention is given in Appendix \ref{color.app}.

After electroweak symmetry breaking, the states in Eq.~(\ref{eq:33}) mix and reclassify themselves according to
color ($\rep3,\ \bar{\rep6}$) and electric charges ($-4/3,\ 2/3,\ -1/3$), denoted by $E_{N_D},U_{N_D},D_{N_D}$.
The relevant interactions among the physical states are then
\begin{eqnarray}
\mathcal{L}_{qqD} &=& K^j_{ab} \left[ \lambda^{E}_{\alpha\beta}
E^j_{N_D}\ \overline{u^C_{\alpha a}}P_\tau u_{\beta b}
+\lambda^{U}_{\alpha\beta} U^j_{N_D}\ \overline{d^{C}_{\alpha a}}P_\tau d_{\beta b}
+\lambda^{D}_{\alpha\beta} D^j_{N_D}\ \overline{d^C_{\alpha b}}P_\tau u_{\alpha a} \right. \nonumber\\
&&
+\lambda^{E'}_{\alpha\beta} E^{j\mu}_{N_D}\  \overline{u^C_{\alpha a}}\gamma_\mu P_R u_{\beta b}
+ \lambda^{U'}_{\alpha\beta}  U^{j\mu}_{N_D}\ \overline{d^C_{\alpha a}}\gamma_\mu P_R d_{\beta b}   \nonumber\\
&&
\left.
+\lambda^{D'}_{\alpha\beta}\ D^{j\mu}_{N_D} \overline{u^C_{\alpha a}}\gamma_\mu P_\tau d_{\beta b} 
\right]
+\rm{h.c.}
\label{eq:qq}
\end{eqnarray}
where $P_{\tau}={1\over2}(1\pm\gamma_5)$ with $\tau=R,L$ for the right- and left-chirality projection operators.
Here and henceforth, we include a superscript $\mu$ to indicate a vector state. 

Naively, the strength of these Yukawa interactions can be naturally of the order of unity,
since the interactions among
colored states are likely to be similar to QCD strong interaction with a coupling constant
$g_{s}^{2} = 4\pi \alpha_{s}\sim {\cal O}(1)$.
However,  many of them are tightly constrained by flavor physics.
A commonly adopted solution is the ``minimal flavor violation'' (MFV) \cite{Chivukula:1987py}.
This assumption makes the couplings align with the SM Yukawa matrices, and they only become
significant when involving heavier quarks such as the top \cite{Arnold:2009ay}.
In some specific model realizations, the MFV is not necessary and certain individual operators involving light
flavors  can be sizable \cite{Mohapatra:2007af}. We do not introduce additional couplings for those new colored
states and thus the baryon number is conserved. In fact, the baryon number can be made a conserved quantum
number for the above interactions by the SM gauge symmetry along with a simple 
extension to the lepton sector  \cite{Arnold:2009ay}.

We note that  the color-triplet scalars ($U_{3},\ D_{3}$)  resemble scalar quarks ($\tilde u,\ \tilde d$) 
in SUSY and the interactions (with the chirality $\tau=L$) are directly analogous to R-parity violating operators
of the $\lambda''$ terms \cite{Barbier:2004ez}, or the ``diquarks''  \cite{Hewett:1988xc}.
Color-triplet scalars at the TeV scale have also been considered in SUSY models to present a unified explanation of dark matter and baryogenesis \cite{Babu:2006wz}.
The color-sextet scalars posses similar nature of ``diquark Higgs'' 
in some unified theories \cite{Pati:1974yy} or some exotic diquarks \cite{Atag:1998xq}.  
The vector states,  on the other hand, are more exotic in terms of connections with an underlying model.
There has been previous interest in the resonant production of diquark scalars and 
vectors at the LHC \cite{Cakir:2005iw}.

\subsection{$\rep3\otimes\rep8$}
\label{38.SEC}
A gluon and a quark can yield large partonic luminosity, and may couple to exotic fermion states.
For simplicity, we only consider spin-$1/2$ states, with quantum numbers as
\begin{eqnarray}
({\rep{3}}\oplus\bar{\rep6},\ \rep{1} \oplus \rep{2})^{1/2}_{-1/3,\ 2/3}.
\nonumber
\end{eqnarray}
 We have not included the $\rep15$ since, as mentioned previously in this section, a $\rep15$
 fermion would spoil the asymptotic freedom of the strong coupling.

Instead of writing down the complete SM gauge invariant operators, 
we consider the interactions after electroweak symmetry
breaking with physical mass eigenstates. These two states are fermionic and  of electric  charges $-1/3$ and $2/3$.
We denote them generically by $q_{N_D}^{*}$, or specifically by $d^*_{N_D}$ and $u^*_{N_D}$, where $N_D=3$ or $6$
for the dimension of their color representation.
The $SU(3)_{C}$ gauge invariance requires the interactions to start with dimension-five, 
and are of the color-magnetic dipole form. 
The Lagrangian for these physical states is then
\begin{eqnarray}
\nonumber
\mathcal{L}_{qgF} =  \frac{g_{s}}{\Lambda}F^{A,\mu\nu}
\hskip -0.2in
&& ~\left[
\bar u {\bar{K}_{N_D,A}} (\lambda^U_LP_L+\lambda^U_RP_R) \sigma_{\mu\nu} u_{N_D}^{*} \right. \\
&& \left. + \bar d {\bar{K}_{N_D,A}} (\lambda^D_LP_L+\lambda^D_RP_R) \sigma_{\mu\nu} d_{N_D}^{*} \right] + \rm{h.c.}
\label{eq:qstar}
\end{eqnarray}
where $F^{A,\mu\nu}$ is the gluon field strength tensor with the adjoint color index $A=1,...,8$,
and ${\bar{K}_A}  $ are $3\times N_D$
matrices of Clebsch-Gordan coefficients connecting the color indices of the different representations.
If the new fermion field is a  ${\rep3}$, then ${\bar{K}_A}^\dagger=K^A=\sqrt{2}T^A$,
where $T^A$ are the fundamental $SU(3)$ representation matrices. Due to the presence of a gluon field,
we naturally include a QCD coupling $g_{s}$.
The new physics scale $\Lambda$ is at least $M_{q^{*}_{j}}$ or higher. In a strongly interacting theory,
we expect that the strength of the couplings $\lambda^U_{L,R}$ and  $\lambda^D_{L,R}$ 
should be typically of the order of unity.
However, if the operators are from one-loop contributions in a weakly coupled theory,
then one would expect to have a suppression factor of the order $1/16\pi^{2}$  \cite{Bauer:2009cc}.

The color-triplet states resemble the excited quarks. 
They could also be string excitations in a low scale string 
scenarios \cite{Cullen:2000ef,Burikham:2004su,Anchordoqui:2008hi}. 
The color-sextet fermions arise in theories of chiral color \cite{Frampton:1987dn}
and top quark condensate models \cite{Martin:1992aq}, the so-called ``quixes''.  There has been previous interest in the color-sextet fermion production at hadron colliders~\cite{Celikel:1998dj}.

\subsection{$\rep8\otimes\rep8$}
\label{88.SEC}
LHC is often referred to as a ``gluon machine" since it has a large parton luminosity for gluon-gluon
initial states. Among the bosonic states from the $\rep8\otimes\rep8$ decompositions,
many higher dimensional color representations can be embedded into  larger theories,
unlike the $\rep15$-tet fermion states that spoil the asymptotic freedom.
We only focus on the color-octet resonances that can result from gluon-gluon fusion.  They may carry the
quantum numbers
\begin{equation}
 ( {\rep8}_{S} \oplus {\rep8}_{A},\ \rep{1} )^{0,1,2}_{0} .
\end{equation}
The symmetric and antisymmetric representations can be utilized with the algebraic relations
of the fundamental representation matrices
\begin{eqnarray}
[ T^A,T^B ]= i f^{ABC}T^C, \quad
\{T^A,T^B\}=\frac{1}{N_C}\delta^{AB}+d^{ABC}T^C,
\end{eqnarray}
and $N_C=3$ is the dimension of the $SU(3)_{C}$ fundamental representation.

The leading operators start from  dimension-five.
Two possible interactions between gluons and a spin zero octet and spin two octet are
\begin{eqnarray}
\mathcal{L}_{gg8}=g_sd^{ABC}\bigg{(}\frac{\kappa_S}{\Lambda_S} S_{8}^{A} F^B_{\mu\nu}F^{C,\mu\nu}+\frac{\kappa_{T}}{\Lambda_T}({T^{A,\mu\sigma}_8}F^{B}_{\mu\nu}{F^{C}_{\sigma}}^{~\nu}+
f {T_{8\ \rho}^{A,\rho}} \
F^{B,\mu\nu}F^C_{\mu\nu})\bigg{)},
\label{tensscal.EQ}
\end{eqnarray}
where $S_8$ ($T_8$) is a scalar (tensor) octet. 
We again assume that the couplings $\kappa_{S}, \kappa_{T}$ of the order
of unity. The relative coupling factor $f$ is more likely to be 1. 
If the operators are from one-loop contributions in a weakly coupled theory,
then one would expect to have a suppression factor of the order $1/16\pi^{2}$.

It is also possible to couple two gluons and a CP-odd octet scalar or tensor.
The couplings of the CP odd states are identical
in form to those in Eq.~(\ref{tensscal.EQ}) with the replacement of one field strength tensor with its dual:
\begin{eqnarray}
\widetilde{F}^{A}_{\mu\nu}=\frac{1}{2}\varepsilon^{\mu\nu\rho\sigma}F^A_{\rho\sigma},
\end{eqnarray}
where $\varepsilon^{\mu\nu\rho\sigma}$ is the four dimensional antisymmetric tensor. 

Finally, the antisymmetric structure constants $f^{ABC}$ can also be used to construct interactions
with CP-odd  color octets.  However, since the color structure is antisymmetric, the Lorentz structure must also be antisymmetric.  Hence, terms proportional to $F_{\mu\nu}\widetilde{F}^{\mu\nu}$ are zero and the only surviving term is $\widetilde{T}_8^{\mu\sigma}\widetilde{F}_{\mu\nu}F^{~\nu}_\sigma$, where $\widetilde{T}_8$ is the CP-odd color-octet tensor.

Color-octet (pseudo)scalars can arise in technicolor models \cite{Hill:2002ap, Eichten:1986eq},
and in universal extra dimensions \cite{Dobrescu:2007xf}.  There has been much recent interest in the gluon fusion production of color-octet scalars at the LHC \cite{Gresham:2007ri,Idilbi:2009cc}.  
These interactions were induced via loops which are parameterized by the octet-scalar coupling 
in Eq.~(\ref{tensscal.EQ}) with an additional suppression from the loop factor. 
Color-octet vector states have also been studied in the context of 
low-scale string resonances \cite{Burikham:2004su,Anchordoqui:2008hi} via gluon-gluon annihilation, but it typically leads to a suppressed rate.


\subsection{$\rep3\otimes\bar{\mathbf{3}}$}
\label{33b.SEC}
Although the quark-antiquark annihilation would not result in the largest partonic luminosity at high energies
in $pp$ collisions, we include some discussions for resonant production
from $\rep3\otimes\bar{\mathbf{3}}$ for completeness.
The resonances may carry the quantum numbers
\begin{equation}
( {\rep1} \oplus {\rep8},\ \rep{1}\oplus \rep{3} )^{1}_{-1,0,1},\quad
( {\rep1} \oplus {\rep8},\ \rep{2} )^{0}_{0,1}.
\end{equation}
Once again, we focus on the color-octet states and ignore the well-known color-singlet states
such as $Z'$'s, $W'$'s, and Kaluza-Klein gravitons.

We first consider the color-octet vector states.
We denote them according to their color and electric charges $V_8^{0},\ V_8^{\pm}$.
We write their interactions with quarks as
\begin{eqnarray}
\mathcal{L}_{q\bar{q}V}&=&g_s\left[ {V_8}^{0,A,\mu}\ \bar{u} T^A \gamma_\mu (g^U_L P_L+g^U_{R}P_R)u+
{V_8}^{0,A,\mu}\ \bar{d} T^A \gamma_\mu (g^D_L P_L+g^D_{R}P_R)d \right .\nonumber\ \\
&&\left .+\left(V_8^{+,A,\mu}\ \bar{u} T^A \gamma_\mu (C_L V^{CKM}_LP_L+C_R V^{CKM}_R P_{R})d+\rm{h.c.}\right)\right ] ,
\label{qqV.EQ}
\end{eqnarray}
where $V^{CKM}_{L,R}$ are the left- and right-handed CKM matrices.
Due to the stringent constraints from flavor physics, we have assumed that
there is no FCNC, and the charged current aligns with the SM CKM.
The couplings $C_{L,R}$ and $g_{L,R}$ are thus diagonal and naturally order of unity.
Well-known examples of color-octet vectors coupled to $q\bar q$ include the
axigluon \cite{Frampton:1987dn,Bagger:1987fz}, a coloron or Techni-$\rho$ \cite{Hill:1991at,Chivukula:1996yr},
a Kaluza-Klein gluon \cite{Agashe:2006hk}, 
and low-scale string resonances \cite{Cullen:2000ef,Burikham:2004su,Anchordoqui:2008hi} via  $q\bar q$ annihilation.

As for the color-octet scalar states, we note that
the renormalizable interactions between a color-octet scalar and two quarks are Yukawa type interactions,
and the SM gauge invariant interactions require the scalar to be a doublet
\cite{Manohar:2006ga} under $SU(2)_{L}$.
Once again, due to the assumption of MFV, their couplings to light quarks would be small, and the only significant coupling
would be to the top or bottom quarks.  The single production of charged and neutral scalars through initial state bottom quarks has been studied previously \cite{FileviezPerez:2008ib}.
However, similar to Higgs production, the dominant resonant production of the scalar states
would be via gluon fusion through top quark loops due to the increased parton luminosity and enhanced couplings.  The effective couplings should be of the same form
as  in Eq.~(\ref{tensscal.EQ}) for $S_{8}$, but with a suppressed coupling.

We summarize the resonant states of our phenomenological interests in Table \ref{qnum.tab2}.
We propose notations for their names, give their conserved quantum numbers, leading couplings to initial state partons,
and related theoretical models.

\begin{table}[tb]
\caption{Summary for resonant particle names, their quantum numbers, and possible underlying models.}
\begin{center}
\begin{tabular}{|c|c|c|c|c|c|}  \hline
 Particle Names & $J$  & $SU(3)_{C}$  & $|Q_{e}|$ & $B$ & Related models \\
 (leading coupling) &  &  &  &  &  \\ \hline
$E_{3,6}^{\mu}\ (uu)$       &0, 1& ${\rep3},\ \overline{\rep6}$ & $4\over3$ & $ -{2\over3} $ & scalar/vector diquarks \\ \hline
$D_{3,6}^{\mu}\ (ud)$       &0, 1& ${\rep3}, \ \overline{\rep6}$ & $1\over3$ & $ -{2\over3} $ & scalar/vector diquarks;
${\tilde d}$ \\  \hline
$U_{3,6}^{\mu}\ (dd)$       &0, 1& ${\rep3},\ \overline{\rep6}$ & $2\over3$ & $ -{2\over3} $ & scalar/vector diquarks;
$\tilde u$ \\  \hline\hline
 $u^{*}_{3,6}\ (ug)$ &$1\over2$, $3\over2$& $\rep3,\ \bar{\rep6}$ & ${2\over 3}$ & $ {1\over3} $ & excited $u$;
 quixes; stringy \\ \hline
  $d^{*}_{3,6}\ (dg)$ &$1\over2$, $3\over2$& $\rep3,\ \bar{\rep6}$ & ${1\over 3}$ & $ {1\over3} $ & excited $d$;
 quixes; stringy \\  \hline\hline
 $S_{8}\ (gg)$   &0& $\rep8_{S}$ &  $0$ & $0$ & $\pi_{TC},\ \eta_{TC} $ \\ \hline
  $T_{8}\ (gg)$   &2& $\rep8_{S}$ &  $0$ & $0$ & stringy \\ \hline \hline
  $V^{0}_{8}\ (u\bar u,\ d\bar d)$      &1& $\rep8$          &  $0$   & $ 0 $  & axigluon; $g^{}_{KK},\ \rho_{TC}$; coloron \\ \hline
  $V_8^{\pm}\ (u\bar d)$  &1& $\rep8$          &  $1$   & $ 0 $  & $\rho^{\pm}_{TC}$; coloron \\ \hline
\end{tabular}
\end{center}
\label{qnum.tab2}
\end{table}


\section{Resonance Production at The LHC}
\label{sec:prodrate}

We will now give analytical formulas and present the expected numerical values of the production cross sections
of the colored resonances at the LHC with 7 and 14 TeV hadronic center of momentum (c.m.)~energies.
The hadronic cross sections are computed by calculating the partonic cross section ($\sigma_{ij}$)
and convolving it  with the parton distribution functions (pdfs).  We write the formula as
\begin{eqnarray}
&& \sigma(S) = \sum_{ij} \int d\tau\ {dL_{ij}\over d\tau}\ \sigma_{ij}(s) , \\
&& {dL_{ij}\over d\tau} \equiv
(f_i\otimes f_j)(\tau)
=\int^1_0 dx_1  \int^1_0 dx_2  f_i(x_1)f_j(x_2) \delta(x_{1}x_{2}-\tau),
\end{eqnarray}
where $S\ (s)$ is hadronic (partonic) c.m.~energy squared, $f_i$ the parton $i$'s
distribution function with a momentum fraction $x_{i}$,
and $\tau=s/S$.
For all numerical results here and henceforth,
we have used the CTEQ6L1~pdfs~\cite{Pumplin:2002vw} and set the factorization and
renormalization scales the same at the resonance mass ($Q^{2}=M^{2}$).

\begin{figure}[tb]
\centering
\subfigure[]{
       \includegraphics[clip,width=0.45\textwidth]{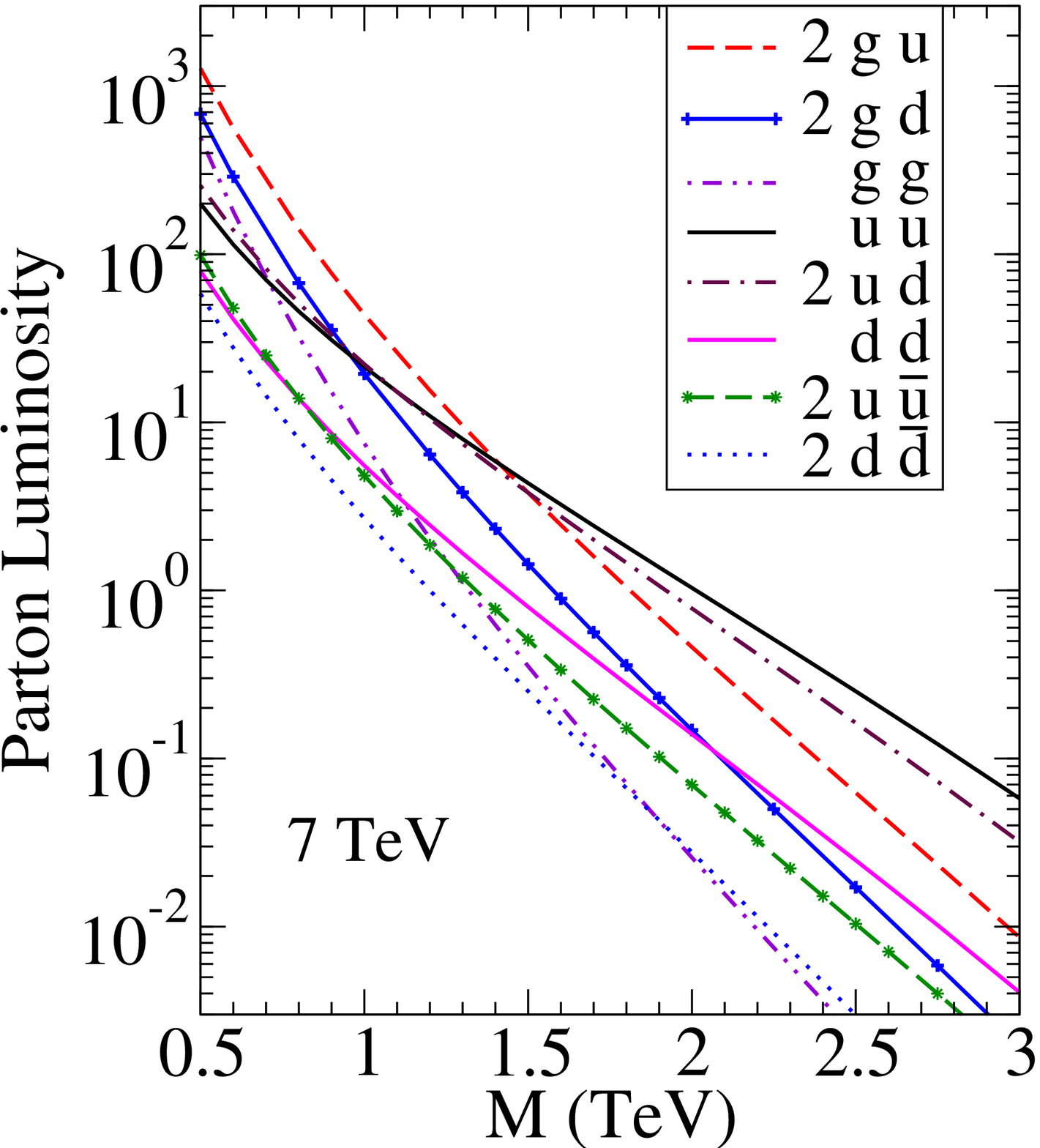}
       \label{7partlum.FIG}
}
\subfigure[]{
       \includegraphics[clip,width=0.45\textwidth]{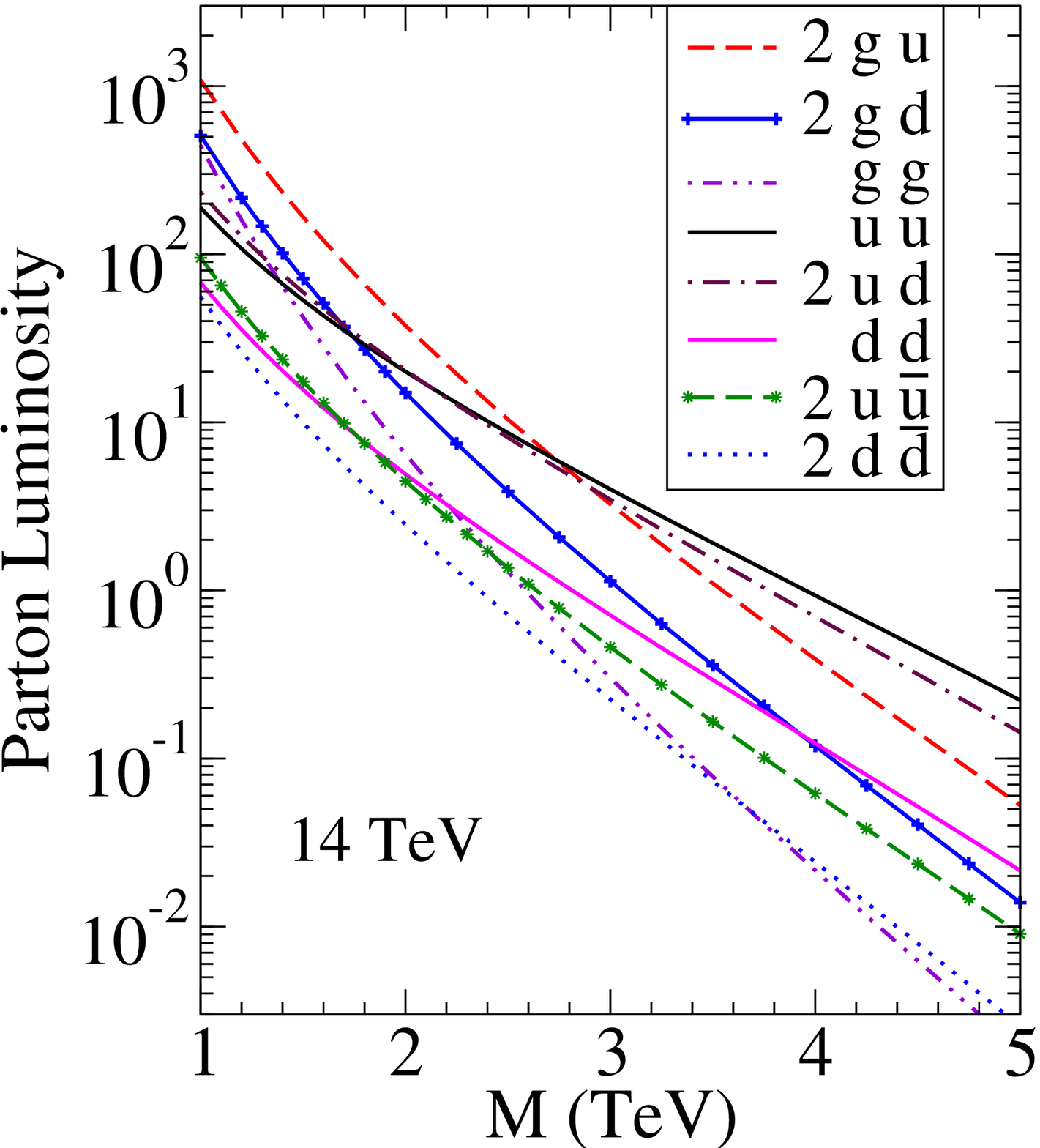}
       \label{14partlum.FIG}
}
\caption{The parton luminosities $dL_{ij} / d\tau$
versus resonance mass at the (a) 7 TeV and (b) 14 TeV LHC.}
\label{partlum.FIG}
\end{figure}

For a resonant production, the on-shell condition forces the partonic cross section to go like
$\sigma_{ij} \sim \delta(s-M^{2})$. Thus the hadronic cross section will be evaluated with
the parton luminosity at $\tau =M^{2}/S$.
We first show the partonic luminosities versus the scale at the resonance mass
in Fig.~\ref{partlum.FIG} for the parton combinations of
$u_{1} u_{2},\ d_{1} d_{2},\ u_{1} d_{2} + d_{1} u_{2},\  g_{1} g_{2,\ }g_{1} u_{2}+u_{1} g_{2},\ g_{1} d_{2} + d_{1} g_{2},\
u_{1} \bar{u}_2 + \bar{u}_1 u_{2},$ and $d_{1}\bar{d}_{2} + \bar{d}_1 d_{2}$
at (a) 7 and (b) 14 TeV LHC.
As expected, initial states involving valence quarks and gluons will have the largest parton luminosities.
In particular, gluons dominate at lower masses, while valence quarks take over at higher masses.
The cross-over between $gg$ and $uu$  occurs near $M=0.75\ (1.2)$ TeV  at  the 7 (14) TeV LHC.
Not only the $u$ quark pdf is about twice as much as that of $d$ at low masses,  but also it falls much
more slowly at high masses than $d$.
For completeness, we also include $\bar q$ initial state when relevant.
In fact, the cross-over of the partonic luminosities  between $gg$ and $u\bar u$  occurs near 
$M=1.2\ (2.2)$ TeV  at  the 7 (14) TeV LHC.

\begin{figure}[tb]
\centering
\subfigure[]{
       \includegraphics[clip,width=0.45\textwidth]{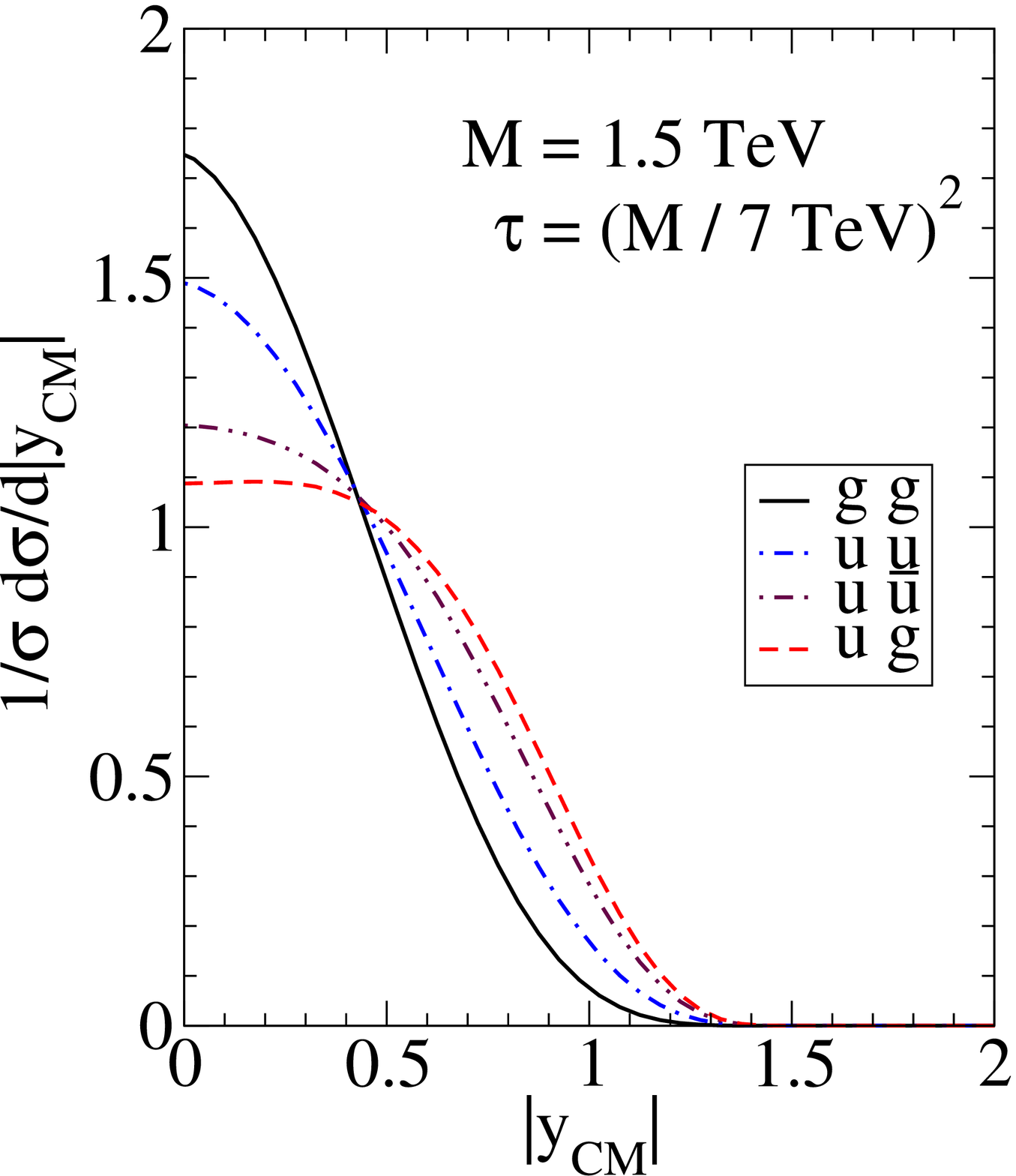}
       \label{7rap.FIG}
}
\subfigure[]{
       \includegraphics[clip,width=0.45\textwidth]{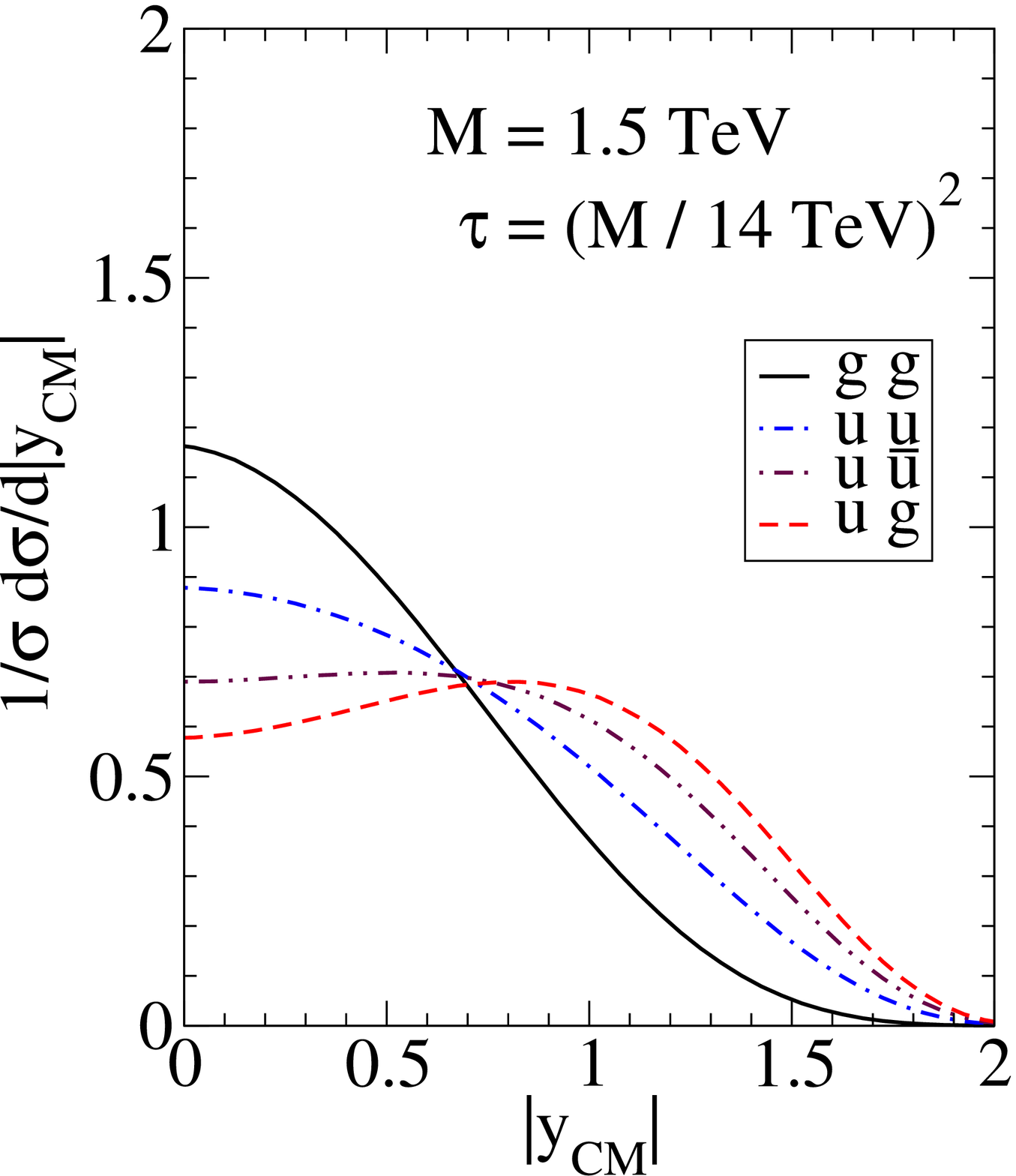}
       \label{14rap.FIG}
}
\caption{Center of momentum system rapidity distributions for resonance mass of $1.5$ TeV and initial states $gg,uu,u\bar{u}$ and $ug$ at the (a) 7 and (b) 14 TeV LHC.}
\label{rap.FIG}
\end{figure}
The rapidity of the partonic c.m.~system  is also of significant interest,  which is defined as
\begin{eqnarray}
y_{CM}^{}=\frac{1}{2}\ln\frac{x_1}{x_2} .
\end{eqnarray}
We show the $y_{CM}^{}$ distributions at the $7$ and $14$ TeV LHC for $M=1.5$~TeV
in Figs.~\ref{7rap.FIG} and \ref{14rap.FIG} respectively.
These distributions measure the longitudinal boost due to the asymmetry between the two parton energies.
The $gg$ and $uu$ initial states are symmetric and hence peaked at zero rapidity.
Due to the up-quark being valence with a broader distribution in $x$,
the $uu$ initial state develops a larger discrepancy between parton momentum fractions than the $gg$ initial state and is therefore broader.  Also, the $u\bar{u},u\bar{d}$ and $ug$ initial states have a large imbalance in the momentum fractions and are broader than the $uu$ and $gg$ initial states. In fact, at the $14$ TeV LHC the imbalance is so pronounced for $ug$ that the rapidity distribution peaks at $|y_{CM}|\approx 0.9$.
Since the $14$ TeV LHC probes lower $\tau$ than the $7$ TeV LHC, a larger discrepancy between the parton momentum fractions can develop and the rapidity distributions are considerably broader than at the $7$ TeV LHC.
This fact will have an impact on the experimental acceptance for the final state jets.

We consider the leading production with the resonances as listed in Table  \ref{qnum.tab2}.
We do not attempt to calculate the decay of the resonances. Instead, we parameterize the production rate to dijets
simply by a branching fraction (BR). Thus the total signal cross section will be governed by a coupling constant to
the initial state partons, a branching fraction, and the resonance mass.

In the following calculations we employ the narrow width approximation, which is valid for $\Gamma \ll M$, where $\Gamma$ and $M$ are the total width and mass of the resonant particle, respectively.  Using the interactions listed in section~\ref{sec:inter}, for a resonance mass on the order of a TeV and order one couplings between the new resonance and SM partons we find $\Gamma\lesssim 0.15\, M$.  However, if the couplings of the resonance are large or there are many additional decay channels, the width may be sizeable and its effects will have to be included.

\subsection{Quark-Quark Annihilation}

\begin{figure}[tb]
\centering
\subfigure[]{
        \includegraphics[clip,width=0.45\textwidth]{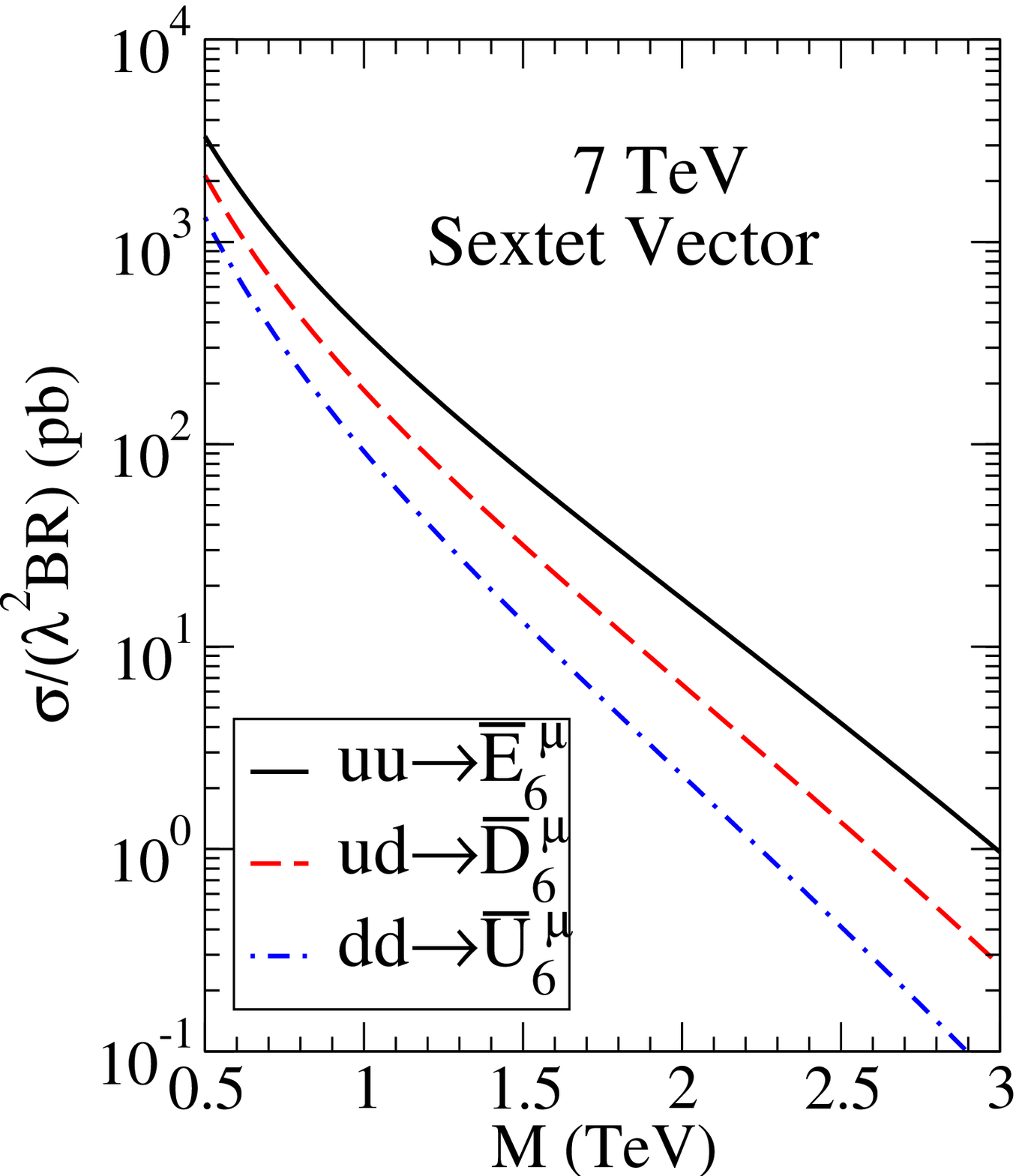}
        \label{7vect6.FIG}
}
\subfigure[]{
        \includegraphics[clip,width=0.45\textwidth]{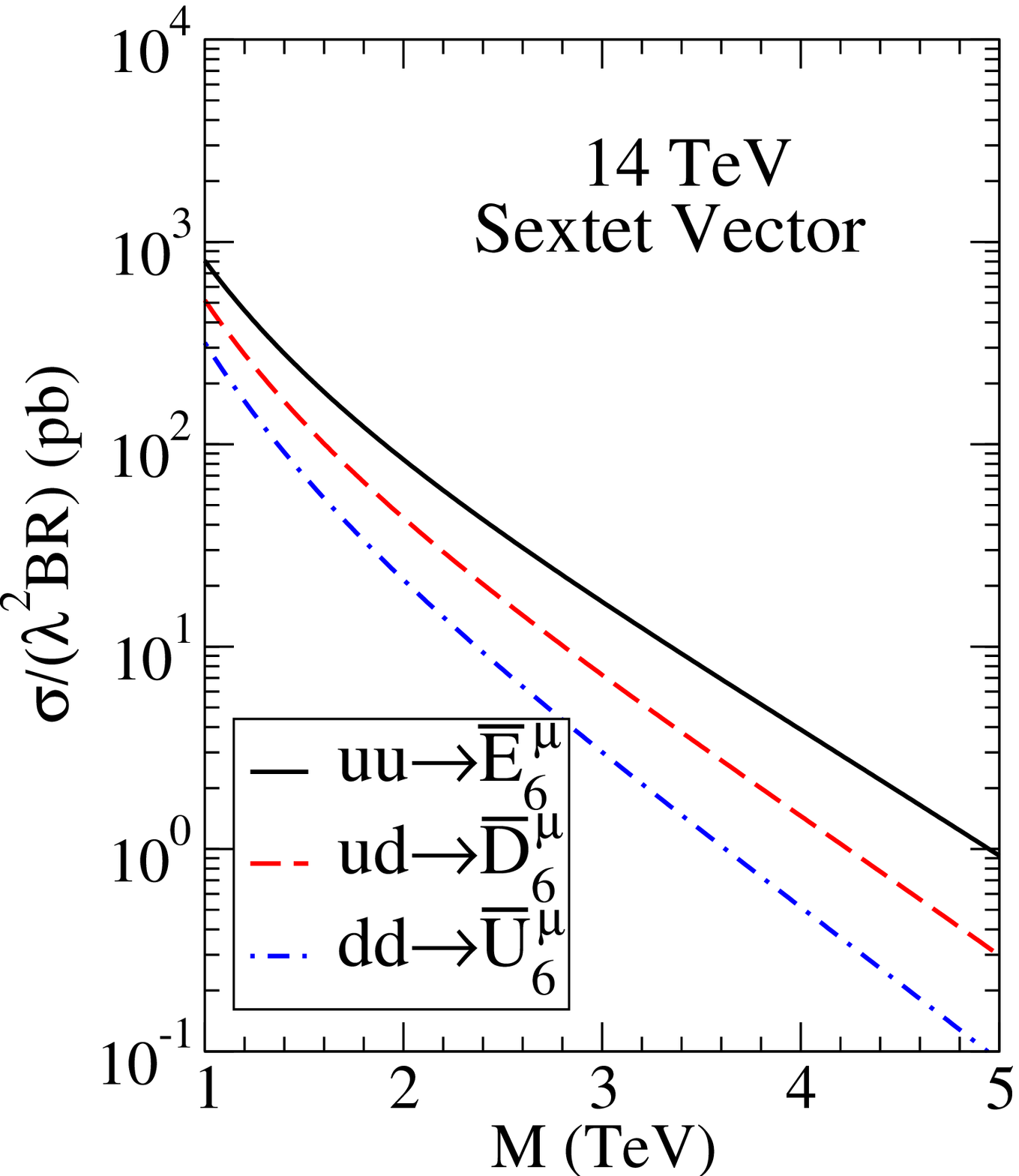}
        \label{14vect6.FIG}
}
\caption{
Dijet cross sections for color-sextet vector production via $uu, ud$ and $dd$ initial states versus its mass
at the LHC for (a) 7 TeV and (b) 14 TeV.
Subleading contributions from antiquarks for the conjugate particle production are also included.
The coupling constant to initial state partons
and the branching fraction to dijet have been factorized out. }
\label{xsect.FIG}
\end{figure}

The $uu$, $dd$, and $ud$ initial states can annihilate into color-antitriplet and sextet
spin 0 and spin 1 particles, often referred as diquarks.  Based on
the interactions of section~\ref{33.SEC} and using the Feynman rules in the appendix, for a resonant
diquark mass of $M$ the hadronic cross section from $uu$ and $dd$ initial states is found to be
%
\begin{equation}
\label{eq:rateDuu}
\sigma_{qq}=\lambda^2\frac{\pi N_D}{N^2_C}\frac{1}{S}(q\otimes q)(\tau_0)
\end{equation}
for both scalar and vector diquarks and for the $ud$ initial state
\begin{equation}
\label{eq:rateD}
\sigma_{ud} = \lambda^2\frac{\pi N_D}{2^{2} N_C^2}\frac{1+\delta_{1J}}{S}\left((u\otimes
d)(\tau_0)+(d\otimes u)(\tau_0)\right),
\end{equation}
%
where the coupling constant  $\lambda$ specifies the resonance as in Eq.~(\ref{eq:qq}) and $J$ is the spin of the resonance.
$N_D$ is the dimension of the antitriplet ($N_D=3$) or sextet ($N_D=6$) representation.
Here and henceforth, $\tau_0=M^2/S$.
%
%

The production cross sections of the color-sextet vector diquarks to dijet
at a 7 TeV and 14 TeV LHC are shown in Figs.~\ref{7vect6.FIG}
and \ref{14vect6.FIG}, respectively.  The production cross sections of the scalars
$E_6,U_6$ are the same as those for the vectors $E^\mu_6,U^\mu_6$ while the production rate for the scalar $D_6$ is half that of $D^\mu_6$.

 Due to the antisymmetric factor on the quark color indices, the
only non-zero valence quark configuration to give a antitriplet scalar diquark is the flavor-off diagonal contribution $ud \to  \bar{D}_3$.
However, the antitriplet vector diquarks can be produced from both the flavor diagonal $uu\rightarrow \bar{E}^\mu_3,\ dd\rightarrow \bar{U}^\mu_3$ and flavor-off diagonal $ud\rightarrow \bar{D}^{\mu}_3$ contributions from valence quarks. 
Also, since the cross section is proportional to the dimension of the diquark representation, the production cross sections for the antitriplet diquarks are half that of the respective sextet diquarks.

Besides the leading contribution from the valence quarks, we have also included the antiquark contributions
for the conjugate particle production in the numerical results presented here.
We summarize a few representative cross sections for the color-sextet vector diquarks,
along with the percentage contribution from the antiquarks.

\begin{equation*}
\begin{array}{llll}
{\rm 7\ TeV\ LHC:}
~~~~~ & \bar E_6^{\mu}~~~~~~~~~~ &\bar D_6^{\mu}~~~~~~~~ & \bar U_6^{\mu} \\
\sigma({\rm pb})\ M=0.5\ {\rm TeV}~~~~~ & 3400    &  2100     &  1300   \\
\bar q \bar q'                          & 2.8\%   &  5.6\%    &  11\%   \\
\sigma({\rm pb})\ M=3\ {\rm TeV}        & 0.96     &  0.27    &  0.064  \\
\bar q \bar q'                          & 0.011\%  &  0.028\% &  0.068\%   \\ \\
{\rm 14\ TeV\ LHC:}
~~~~~ & \bar E_6^{\mu}~~~~~ & \bar D_6^{\mu}~~~~~ & \bar U_6^{\mu} \\
\sigma({\rm pb})\ M=1\ {\rm TeV} & 800     &  510       &  320   \\
\bar q \bar q'                   & 2.8\%   &  5.5\%     &  11\%   \\
\sigma({\rm pb})\ M=5\ {\rm TeV} & 0.92    &  0.30       &  0.090  \\
\bar q \bar q'                   & 0.026\% &  0.075\%    &  0.21\%
\end{array}
\end{equation*}
Once again, we have pulled out the
coupling constant $\lambda^{2}$ and the branching fraction (or equivalent to setting $\lambda^{2}=$BR=1).

The next-to-leading order (NLO) QCD corrections to scalar diquark production have been previously
calculated \cite{Han:2009ya} and sizable corrections were found.
For instance, the cross section  with masses between $0.5$ and $1.5$ TeV can be increased  by about
 $20\%$ for $U_6$ and $30-35\%$ for $D_3$. 
 It is expected that the corrections to the other color-sextets and antitriplets 
should  be similar to the above. 
 %

%


\subsection{Quark-Gluon Annihilation}

\begin{figure}[tb]
\centering
\subfigure[]{
       \includegraphics[clip,width=0.45\textwidth]{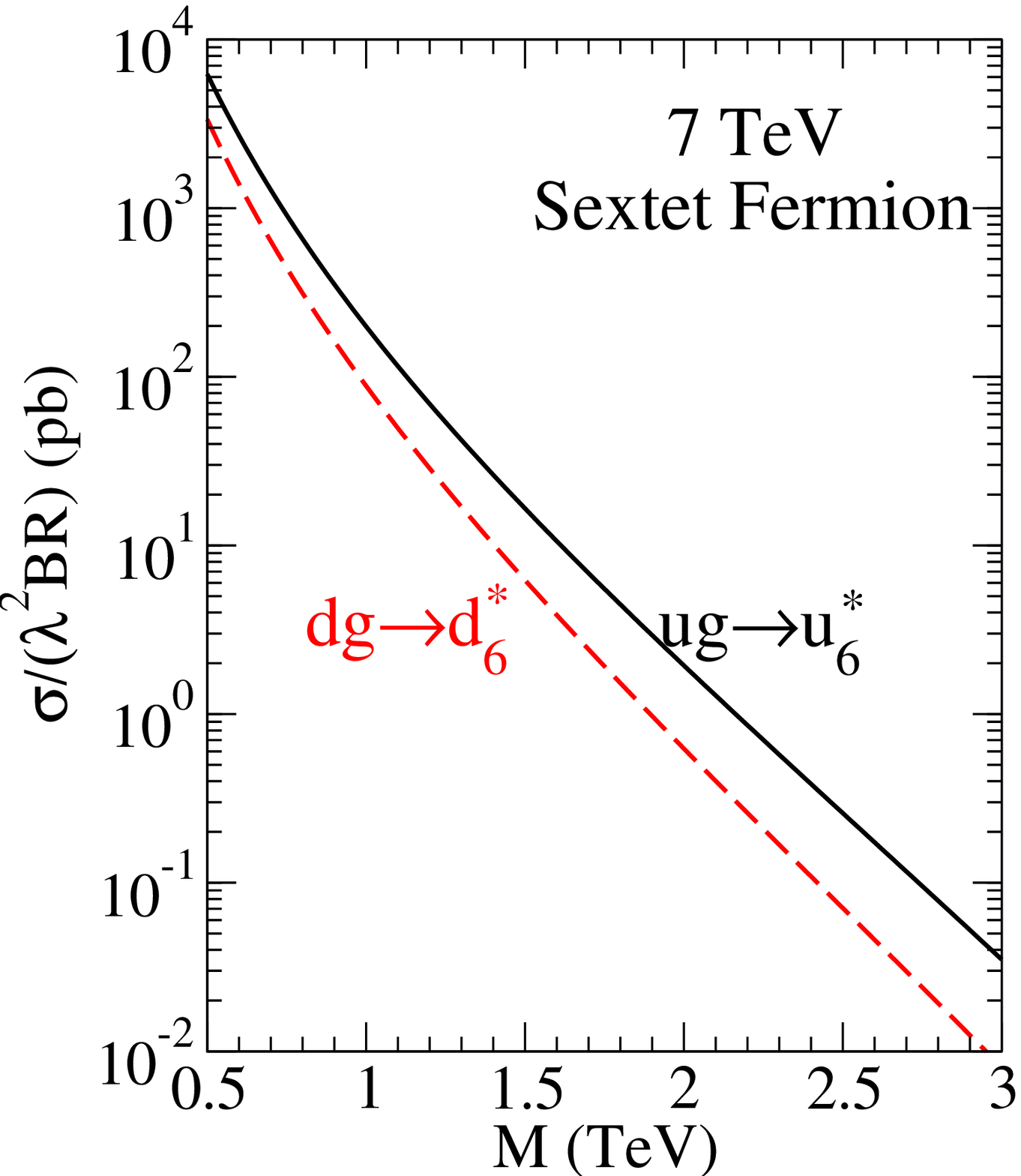}
       \label{7ferm6.FIG}
}
\subfigure[]{
       \includegraphics[clip,width=0.45\textwidth]{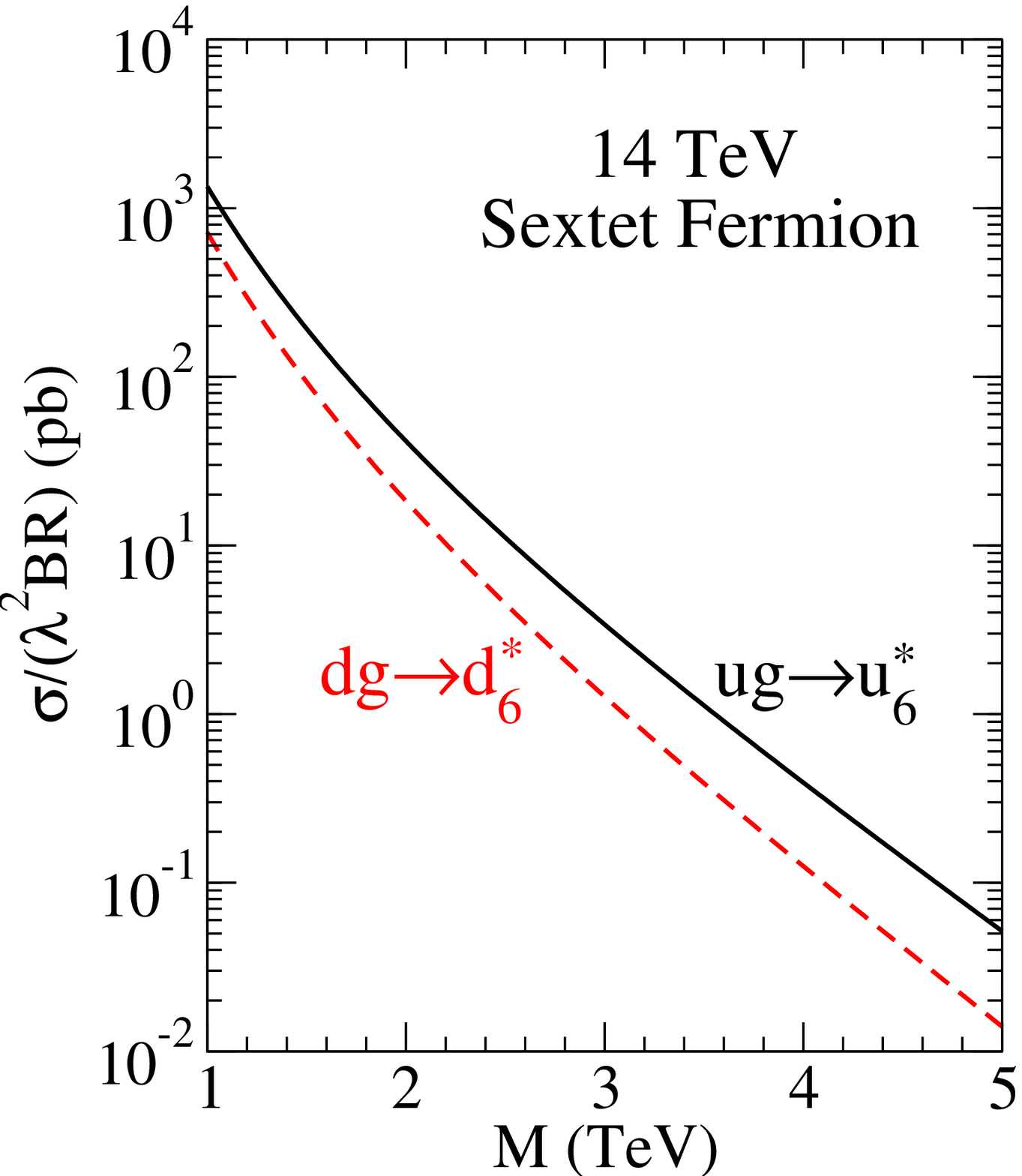}
       \label{14ferm6.FIG}
}
\caption{
Dijet cross sections for color-sextet fermion production via $ug$ and $dg$ initial states versus its mass
at the LHC for (a) 7 TeV and (b) 14 TeV.
Subleading contributions from antiquarks for the conjugate particle production are also included.
The coupling constant to initial state partons
and the branching fraction to dijet have been factorized out.
The new physics scale, $\Lambda$, has been set equal to $2M$.}
\label{fermsext.FIG}
\end{figure}

The $ug$ and $dg$ initial states can produce color-triplet and sextet excited quarks.  Using the Feynman rules in the
appendix, the hadronic cross section for excited quarks of mass $M$ is
\begin{eqnarray}\label{eq:rate6F}
\sigma_{qg}=8\pi^2\lambda^2\frac{\alpha_s}{N_C}\frac{M^2}{\Lambda^2}\frac{1}{S}(g\otimes q)(\tau_0),
\end{eqnarray}
where $\lambda^2={\lambda^{2}_L}+\lambda^2_{R}$ and $\lambda_{L,R}$ specify the interactions
as in Eq.~(\ref{eq:qstar}).
Since the Clebsch-Gordan coefficients for the color-triplet and sextet states are normalized the same it follows that the production cross sections are the same.
Comparing with the convention in Ref.~\cite{Baur:1987ga}, the
new physics scale $\Lambda$ here corresponds to twice the excited quark mass.

Figure~\ref{fermsext.FIG} presents the production cross section of excited sextet
quarks $u^*_6$ and $d^*_6$ produced from $ug$ and $dg$ initial states,
respectively, at the  (a) 7 and (b) 14 TeV LHC.
The $u^*_6$ production rate is larger than the $d^*_6$ production rate by about a factor of two
at low mass, due to  the larger $u$ quark pdf. We have taken the cutoff $\Lambda =2M$ in the
numerical evaluation.  With our CG coefficient normalization, our results should be a factor of two
larger than that using the convention of \cite{Baur:1987ga}.

For the numerical results we have once again included the conjugates, produced
from $\bar{u}g, \bar d g \rightarrow \overline{u}^*, \bar{d}^{*}$.  Representative results for the total cross section
and the percentage contribution from antiquarks, after factorizing out the overall constants, are
\begin{equation*}
\begin{array}{lll}
{\rm 7\ TeV\ LHC:}
~~~~~ &  u_{6}^{*}~~~~~~~~~~ & d_{6}^{*} \\
\sigma({\rm pb})\ M=0.5\ {\rm TeV}~~~~~ & 6200 & 3400   \\
\bar q g                                & 8.5\%     &  20\%    \\
\sigma({\rm pb})\ M=3\ {\rm TeV}        & 0.035 &  0.0080  \\
\bar q g                                & 0.82\%     &  2.6\%       \\ \\
{\rm 14\ TeV\ LHC:}
~~~~~ &  u_{6}^{*}~~~~~ & d_{6}^{*} \\
\sigma({\rm pb})\ M=1\ {\rm TeV}        &  1300      &  720  \\
\bar q g                                &  8.4\%     &  20\%    \\
\sigma({\rm pb})\ M=5\ {\rm TeV}        &  0.052     &  0.014  \\
\bar q g                                &  1.2\%     &  4.2\%
\end{array}
\end{equation*}

\subsection{Gluon-Gluon Annihilation}

Gluon-gluon annihilation can result in color-octet scalars and tensors.
Using the parameterization of Eq.~(\ref{tensscal.EQ}) and Feynman rules in the appendix, the hadronic production cross section of a color-octet scalar and tensor of mass $M$ from gluon-gluon fusion is
\begin{eqnarray}
\sigma_{gg}=4\pi^2\alpha_s\kappa^2\frac{N^2_C-4}{N_C(N^2_C-1)}\frac{M^2}{\Lambda^2}\frac{1+\delta_{0J}}{S}(g\otimes g)(\tau_0),
\end{eqnarray}
where $\kappa$ and $\Lambda$ are specified by the interaction.
Since on-shell tensor polarizations are traceless, the ${{T_8}^\rho}_\rho$ term in Eq.~(\ref{tensscal.EQ}) does not contribute to the resonant production of the color-octet tensor.

\begin{figure}[tb]
\centering
\includegraphics[clip,width=0.45\textwidth]{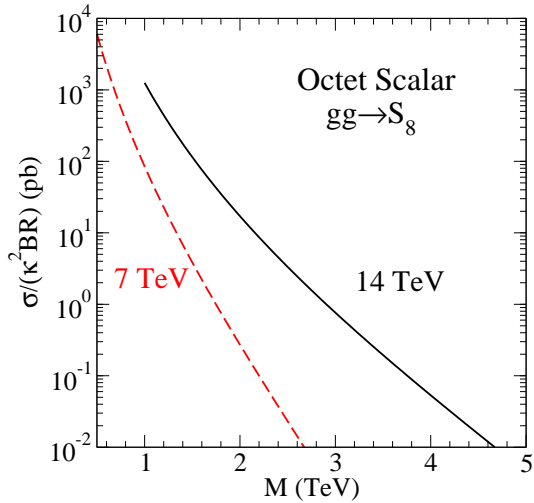}
\caption{Dijet cross sections for color-octet scalar production via $gg$ initial states versus its mass
at the LHC for 7 TeV (dashed curve) and 14 TeV (solid curve).
The coupling constant to initial state partons
and the branching fraction to dijet have been factorized out. The new physics scale, $\Lambda$, has been set equal to $M$.}
\label{oct.FIG}
\end{figure}

The production cross sections for the color-octet scalar to dijets are presented in Fig.~\ref{oct.FIG} for the LHC
at 7 TeV (dashed curve) and 14 TeV (solid curve).
For the numerical results presented the new physics scale $\Lambda$
has been set equal to the resonant mass.
The color-octet tensor cross section is one half that of the the color-octet scalar.
Since the gluon luminosity falls fast at a higher mass, the cross section
at 7 TeV LHC drops by more than five orders of magnitude from 6000 pb at $M=0.5$ TeV to 0.02 pb at $M=2.5$ TeV,
and
at 14 TeV LHC by about five orders of magnitude from 1200 pb at $M=1$ TeV to 0.01 pb at $M=4.6$ TeV.  The production cross section of an octet tensor is half that of the octet scalar.

The next-to-leading logarithm (NLL) soft-gluon resummation correction to scalar octet production via gluon fusion has been previously
calculated and sizable corrections were found~ \cite{Idilbi:2009cc}.   The cross section  can be increased by a factor of $2.4$ at mass $0.5$ TeV and $3.5$ at a mass of $2.5$~TeV.

\subsection{Quark-Antiquark Annihilation}

Although the parton luminosity is lower than the previously discussed initial states, we also include resonant production from $u\bar{u}$, $d\bar{d}$, $u\bar{d}$, and $d\bar{u}$ initial states.  These states can couple to color-octet vectors.
Using the interactions in Eq.~(\ref{qqV.EQ}), the production cross section for a color-octet vector of mass $M$ from $q\bar{q}'$ initial states is
\begin{eqnarray}
\sigma_{q\bar{q}}=4\pi^2 g^2\alpha_s\frac{C_F}{N_C}\frac{1}{S}(q\otimes\bar{q}')(\tau_0)
\label{xsect8.EQ}
\end{eqnarray}
where

\begin{eqnarray}
\nonumber
g^{2} =
\left\{
\begin{array}{ll}
\frac{1}{2}(|C_L V^{CKM}_L|^2+|C_R V^{CKM}_R|^2)   &  \quad  {\rm  for\ charged\ states}, \\
 \frac{1}{2}(|g^{U,D}_L|^2+|g^{U,D}_R|^2) & \quad  {\rm for\ neutral\ states}.
\end{array}
\right.
\end{eqnarray}
The color factor $C_F=(N_{C}^{2}-1)/2N_{C} = 4/3$.

\begin{figure}[tb]
\centering
\subfigure[]{
        \includegraphics[clip,width=0.45\textwidth]{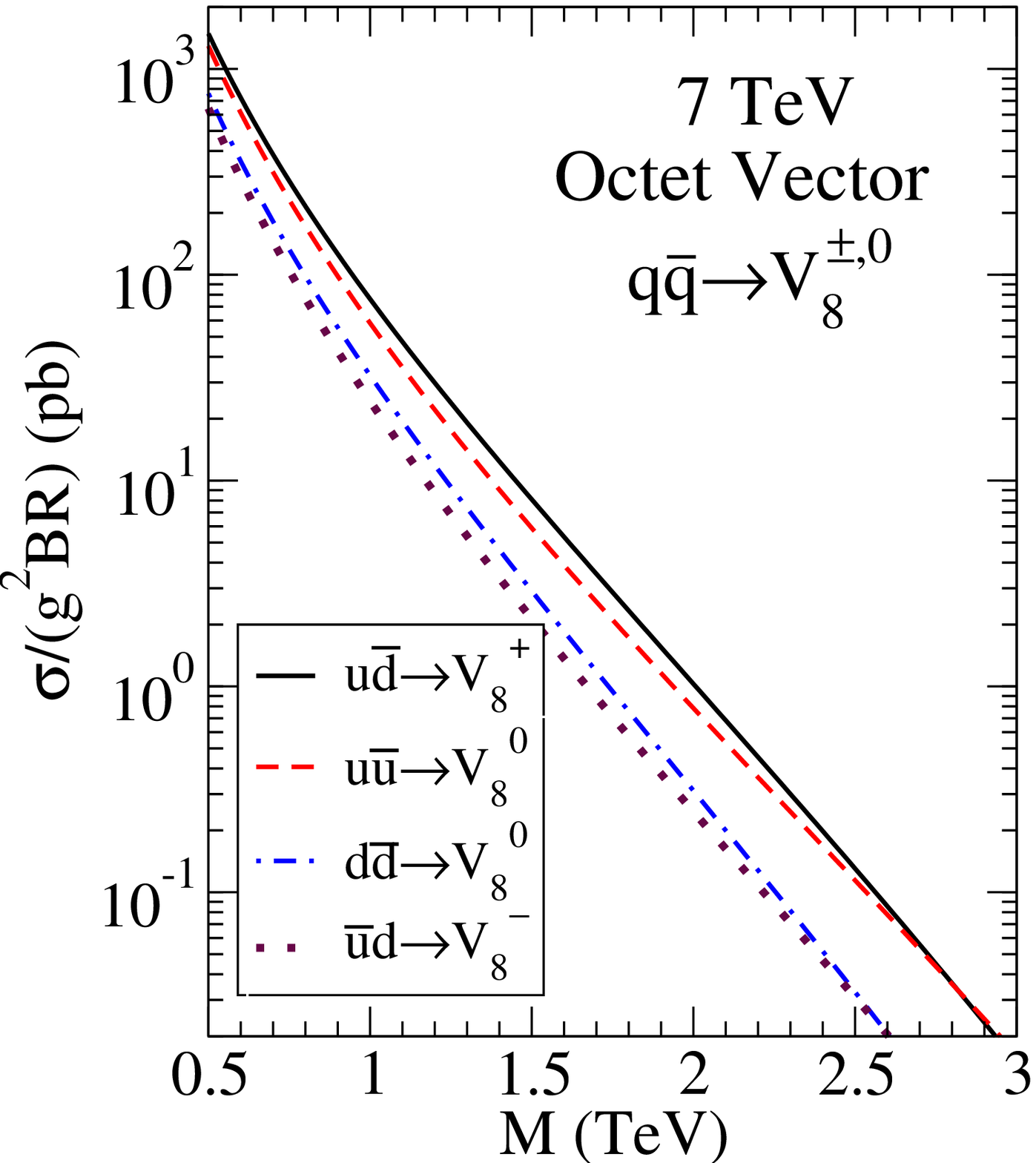}
        \label{7vect81.FIG}
}
\subfigure[]{
        \includegraphics[clip,width=0.45\textwidth]{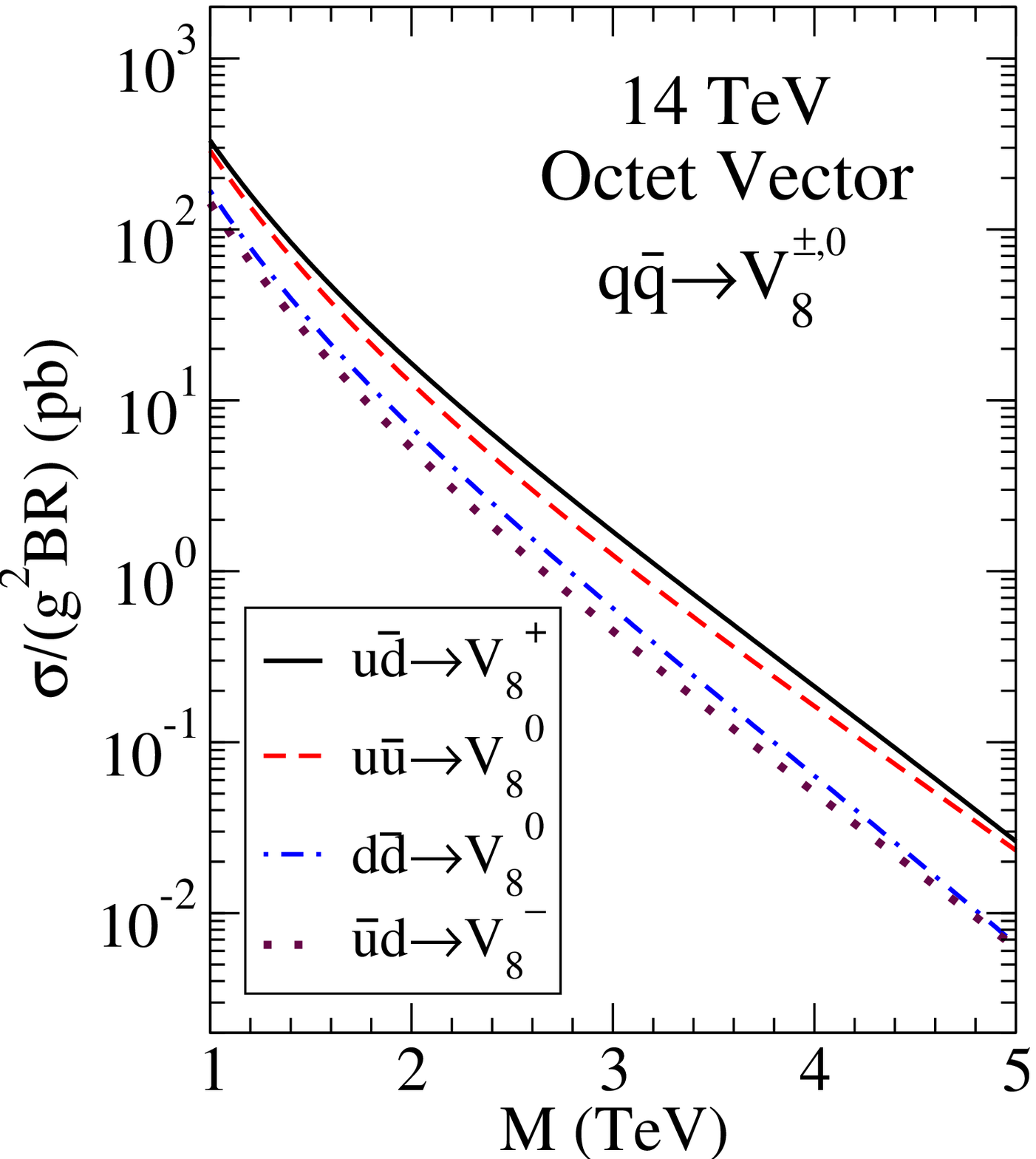}
        \label{14vect81.FIG}
}
\caption{Dijet cross sections for color-octet vector production via $u\bar{d}, u\bar{u}$, $d\bar{d}$, and $d\bar{u}$ initial states versus its mass
at the LHC for (a) 7 TeV and (b) 14 TeV.  The coupling constant to initial state partons
and the branching fraction to dijet have been factorized out.}
\label{xsect81.FIG}
\end{figure}

The cross sections for color-octet vectors are presented in Fig.~\ref{xsect81.FIG} for the (a) 7 TeV and (b) 14 TeV LHC.  
Since the $u$ quark pdf is greater than the $d$ quark pdf, the neutral vectors are produced more favorably by the $u\bar{u}$ initial state than by $d\bar{d}$.  Due to the $\bar{d}$ quark pdf being larger than the $\bar{u}$ quark pdf, the production of $V^+_8$ is larger than the production rate of $V^0_8$ from $u\bar{u}$ initial state and $V^0_8$ production rate from $d\bar{d}$ initial state is larger than the production rate of $V^-_8$. 
Some representative cross sections values are listed below.
\begin{equation*}
\begin{array}{lllll}
{\rm 7\ TeV\ LHC:}
~~~~~ & u\bar d \to V_{8}^{+}~~~~~~~ & d\bar u \to V_{8}^{-}~~~~~~~& u\bar u\to V_{8}^{0}~~~~~~~ & d\bar d \to V_{8}^{0} \\
\sigma({\rm pb})\ M=0.5\ {\rm TeV}~~~~~ & 1500   & 640              & 1300                       &  760 \\
\sigma({\rm pb})\ M=3\ {\rm TeV}        & 0.015  &   0.0037         & 0.016                      & 0.0030  \\ \\
{\rm 14\ TeV\ LHC:}
~~~~~ & u\bar d \to V_{8}^{+}~~~~~~~ & d\bar u \to V_{8}^{-}~~~~~~~ & u\bar u\to V_{8}^{0}~~~~~~~ & d\bar d \to V_{8}^{0} \\
\sigma({\rm pb})\ M=1\ {\rm TeV} & 330    &  140                    & 290                         &170 \\
\sigma({\rm pb})\ M=5\ {\rm TeV} & 0.026   &  0.0063                    &  0.023                      & 0.0066\\
\end{array}
\end{equation*}
Once again, we have pulled out the
coupling constant $\lambda^{2}$ and the branching fraction (or equivalent to setting $\lambda^{2}=$BR=1).


All the cross sections presented in this section are at leading order in QCD.
The production cross section of colored resonance can receive sizable QCD corrections
as shown for the color-triplet and sextet scalar diquarks \cite{Han:2009ya} and color octet scalars~\cite{Idilbi:2009cc}.  We will take this
into account when setting the bounds.

Throughout this paper, we neglect the color-singlet states, such as $Z',\ W'$ or KK gravitons. Our formalism for
is equally applicable to those by adjusting the couplings $g_{s}\to e/\sin\theta_{W}$ and setting the
color factor $C_{F}$ to 1. Before folding in the decay branching fraction to the final state, the production
rates for a color-singlet state would be smaller than the colored resonance by roughly about a factor of 30.


\section{Bounds From the LHC Dijet Spectrum}
\label{sec:bnds}

Searching for new physics signals in the dijet spectrum at hadron colliders has been long carried out.  The Tevatron~\cite{:2009mh} and LHC~\cite{Collaboration:2010ez} experiments have both utilized measurement of the dijet angular distribution to bound the strength of four-fermion contact interactions.
The standard form of the four-fermion contact interaction in the literature is \cite{Eichten:1983hw}:
\begin{equation}
\mathcal{L}_{4q}=\frac{2\pi}{\Lambda^2}\bar{q}_L\gamma^\mu q_L\bar{q}_L\gamma_\mu q_L
\label{4Q.EQ}
\end{equation}
For a sufficiently high mass, the new resonant states can be integrated out and the $\rep3\otimes\rep3$ and $\rep3\otimes\bar{\rep3}$ vector interactions can produce similar interactions to Eq.~(\ref{4Q.EQ}).  The bounds on the four fermion interactions can be roughly translated into bounds on our interactions with the identification
\begin{equation}
\frac{2\pi}{\Lambda^2}\sim \frac{\lambda^2}{2M^2},
\end{equation}
where $M$ is the mass of the resonant state of our current interest.
Assuming a coupling constant of unity, the current LHC bound of $\Lambda\ge4$~TeV translates into
$$M \gtrsim 1.1~{\rm TeV}.$$
Note this bound is only a rough estimate since one would have to be careful in computing the color factor and counting the contributing light partons.

\begin{figure}[tb]
\centering
\subfigure[]{
\includegraphics[clip,width=0.45\textwidth]{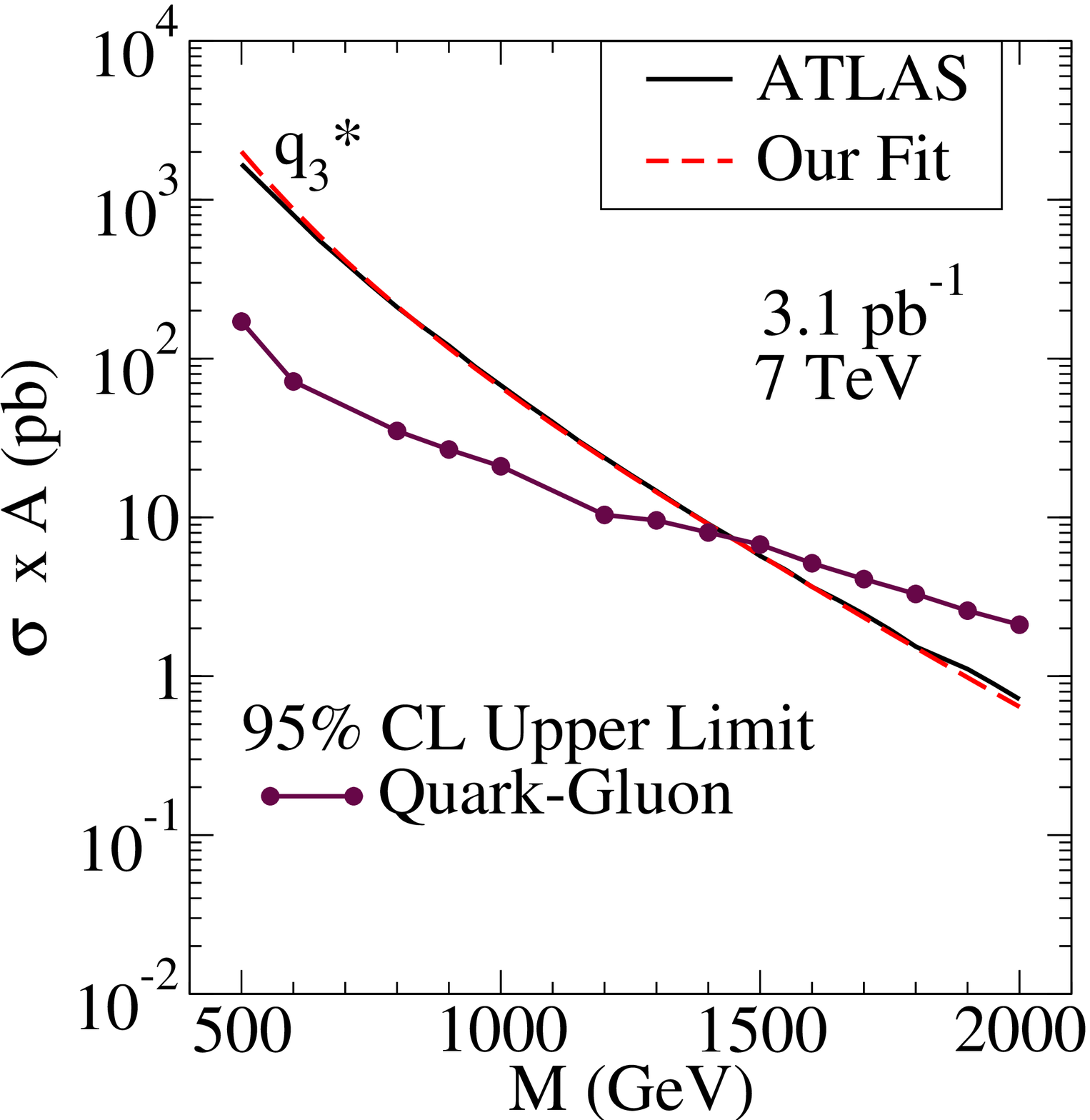}
\label{ATLAS.FIG}
}
\subfigure[]{
\includegraphics[clip,width=0.45\textwidth]{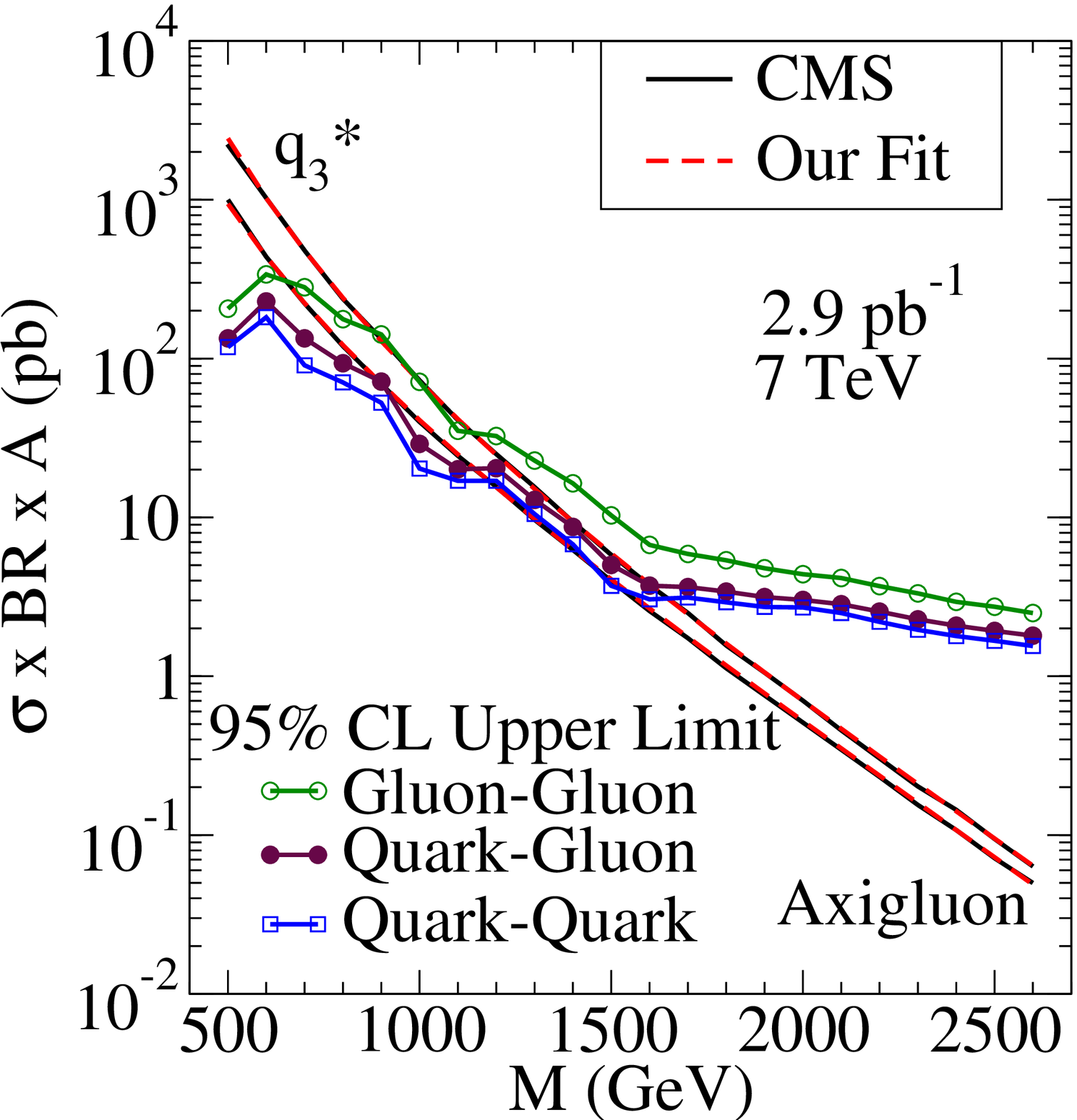}
\label{CMS.FIG}
}
\caption[]{$95\%$ confidence level upper limits on dijet production cross sections versus resonant mass for 
(a) ATLAS results (solid circles) and (b) CMS results from the contributions of 
gluon-gluon (open circles), quark-gluon (solid circles), and quark-quark (open boxes).
Our fits (dashed curve) almost overlap with the theoretical predictions  (solid curves)
provided by  ATLAS for $q^{*}_{3}$ and CMS  for $q^{*}_{3}$  and axigluon. }
\label{dijet.FIG}
\end{figure}

Using measurements of dijet production rates at 7 TeV LHC,
the ATLAS and CMS collaborations have recently released bounds on the dijet production
cross sections as a function of resonance mass based on  the first  data of $3.1$ pb$^{-1}$ \cite{Collaboration:2010bc}
and $2.9$ pb$^{-1}$ \cite{Collaboration:2010jd}, respectively. Even with such small amount of initial data, the LHC experiments have
gone beyond the existing Tevatron results, pushing the LHC to the phase of discovery for new physics.

We model the experimental efficiencies by a simple parameterization. The detector acceptance for dijet events at ATLAS was about $31\%$ for an excited quark mass around $300$ GeV, and about $48\%$ around $1700$ GeV.
For our study we model this acceptance as
\begin{equation}
\mathcal{A}_{ATLAS}=\left\{\begin{array}{l l} \displaystyle\frac{0.17}{1400~\rm{GeV}}(m-300~\rm{GeV})+0.31 & ~~~m\le 1700~\rm{GeV}\vspace{0.1in}\\  0.48 &~~~m> 1700~\rm{GeV}\end{array}\right.
\label{ATLAS.EQ}
\end{equation}
To model the CMS detector acceptances we compared our results for dijet production cross section without detector acceptance to the the CMS results including detector acceptance.
Using their axigluon and excited quark results, we model the CMS detector acceptance as
\begin{equation}
\mathcal{A}_{CMS}=\frac{\Delta}{2100~\rm{GeV}}(m-500~\rm{GeV})+0.47,
\label{CMS.EQ}
\end{equation}
where $\Delta=0.08$ for the quark-quark final state and $\Delta=0.17$ for the quark-gluon final state.  There was no analogous data to find the acceptance for gluon-gluon final states.  We therefore also use the quark-gluon acceptance for the gluon-gluon final state.

The predicted dijet cross sections for triplet excited quarks at ATLAS and triplet excited quarks and axigluons at CMS are presented by the solid curves in Figs.~\ref{ATLAS.FIG} and~\ref{CMS.FIG}, respectively. To reproduce their results for the triplet excited quark production, we set $\lambda=1$ in Eq.~(\ref{eq:rate6F}) and summed over all possible initial state quarks.  As can be seen, using the acceptances in Eqs.~(\ref{ATLAS.EQ}) and~(\ref{CMS.EQ}), our simulations (dashed
curves) fit well the results provided by the ATLAS and CMS collaborations (solid curves).  
The current $95\%$ confidence level upper limits for dijet production cross sections at both ATLAS and CMS are also presented in Fig.~\ref{dijet.FIG}. 

\begin{figure}[tb]
\centering
\subfigure[]{
\includegraphics[clip,width=0.45\textwidth]{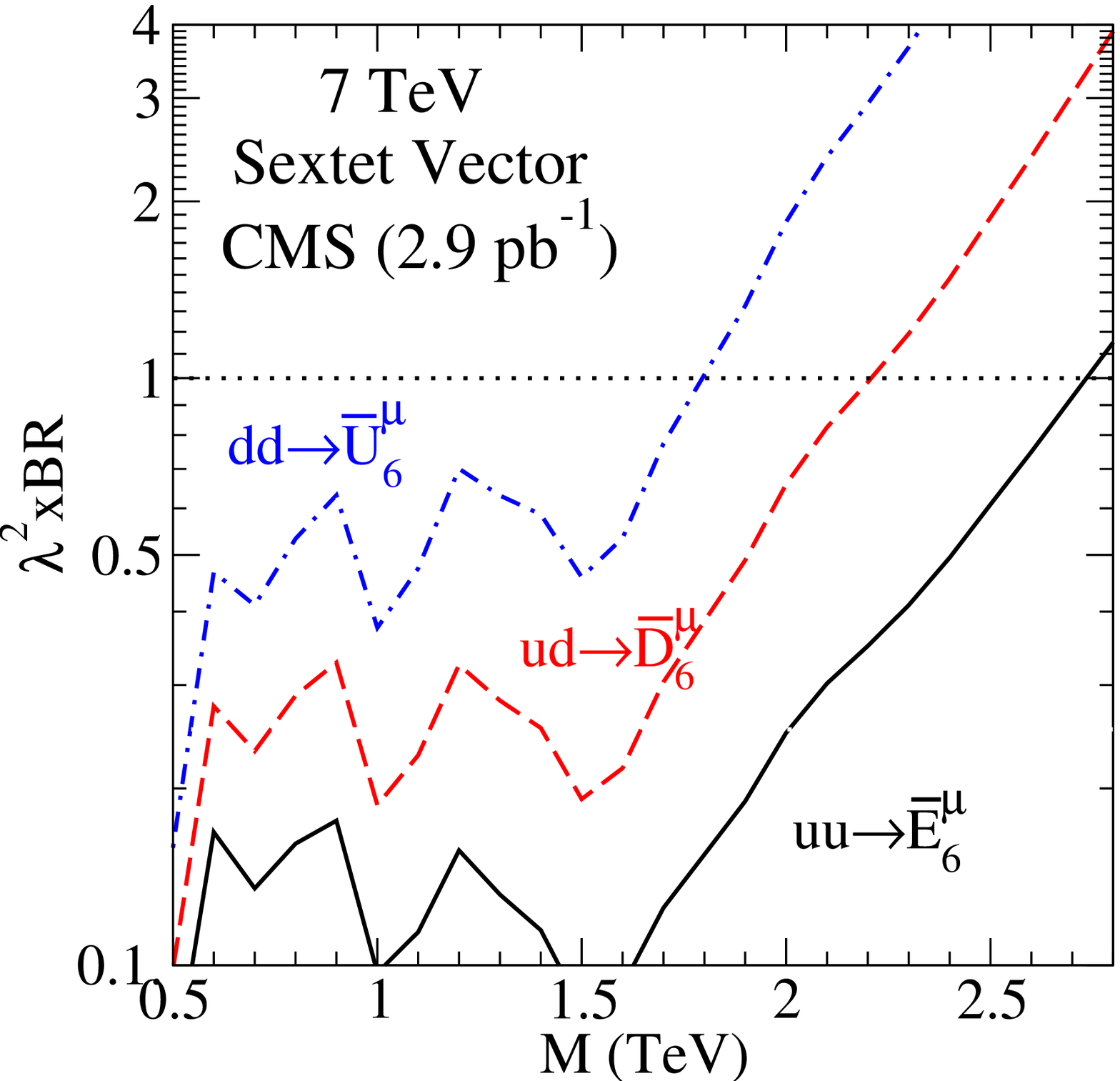}}
\subfigure[]{
\includegraphics[clip,width=0.45\textwidth]{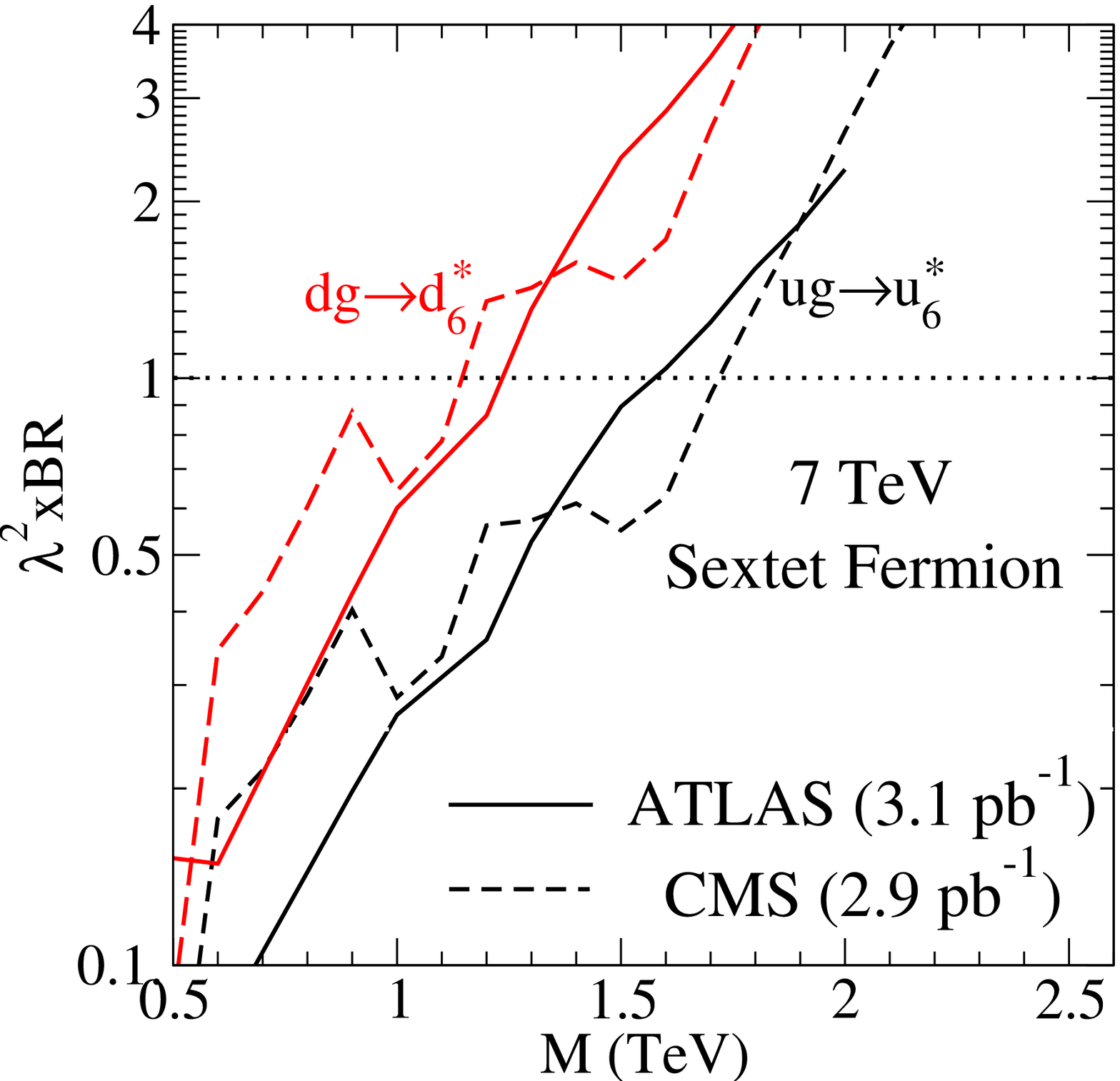}}\\
\vspace{0.01in}
\subfigure[]{
\includegraphics[clip,width=0.45\textwidth]{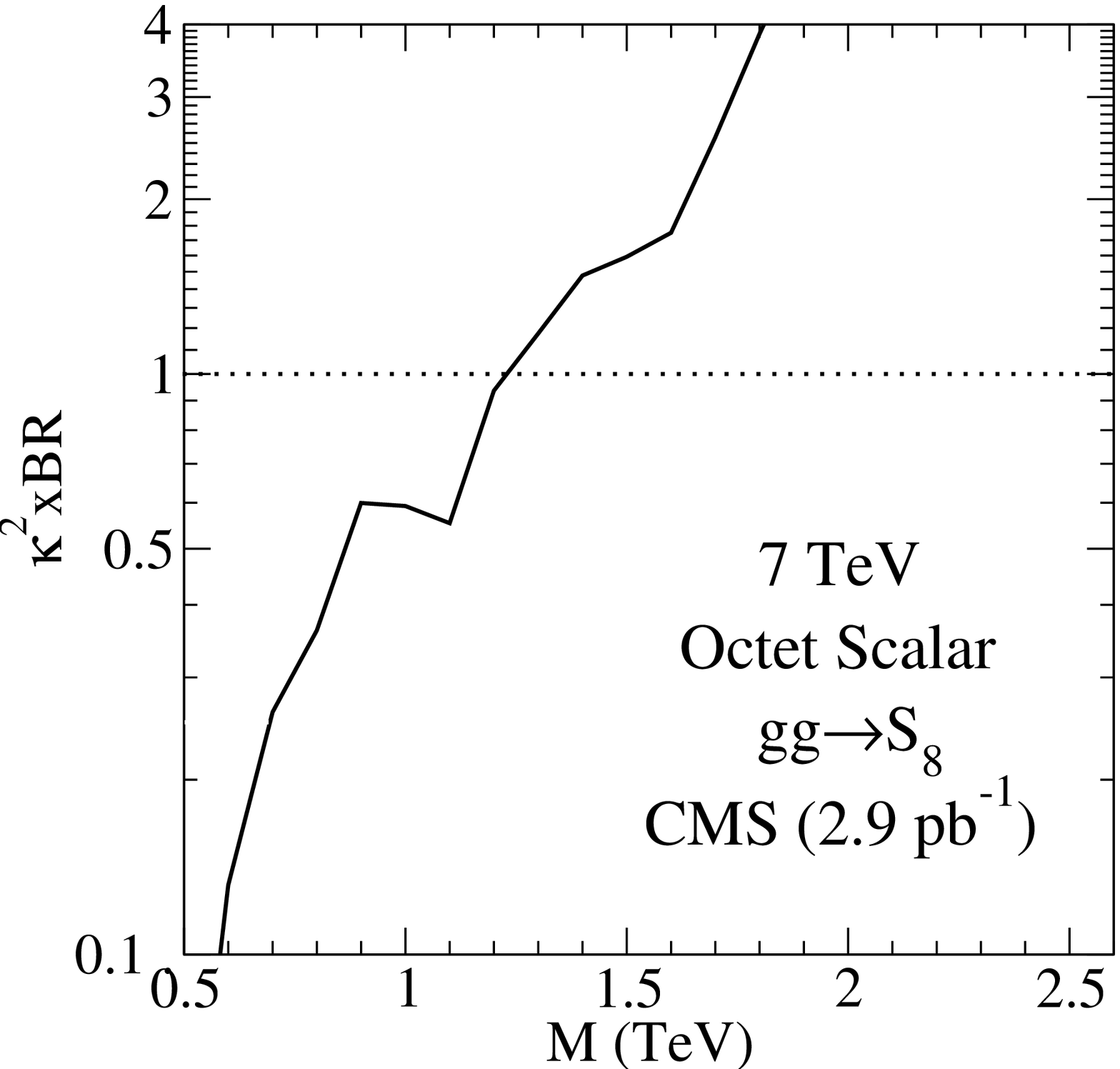}}
\subfigure[]{
        \includegraphics[clip,width=0.45\textwidth]{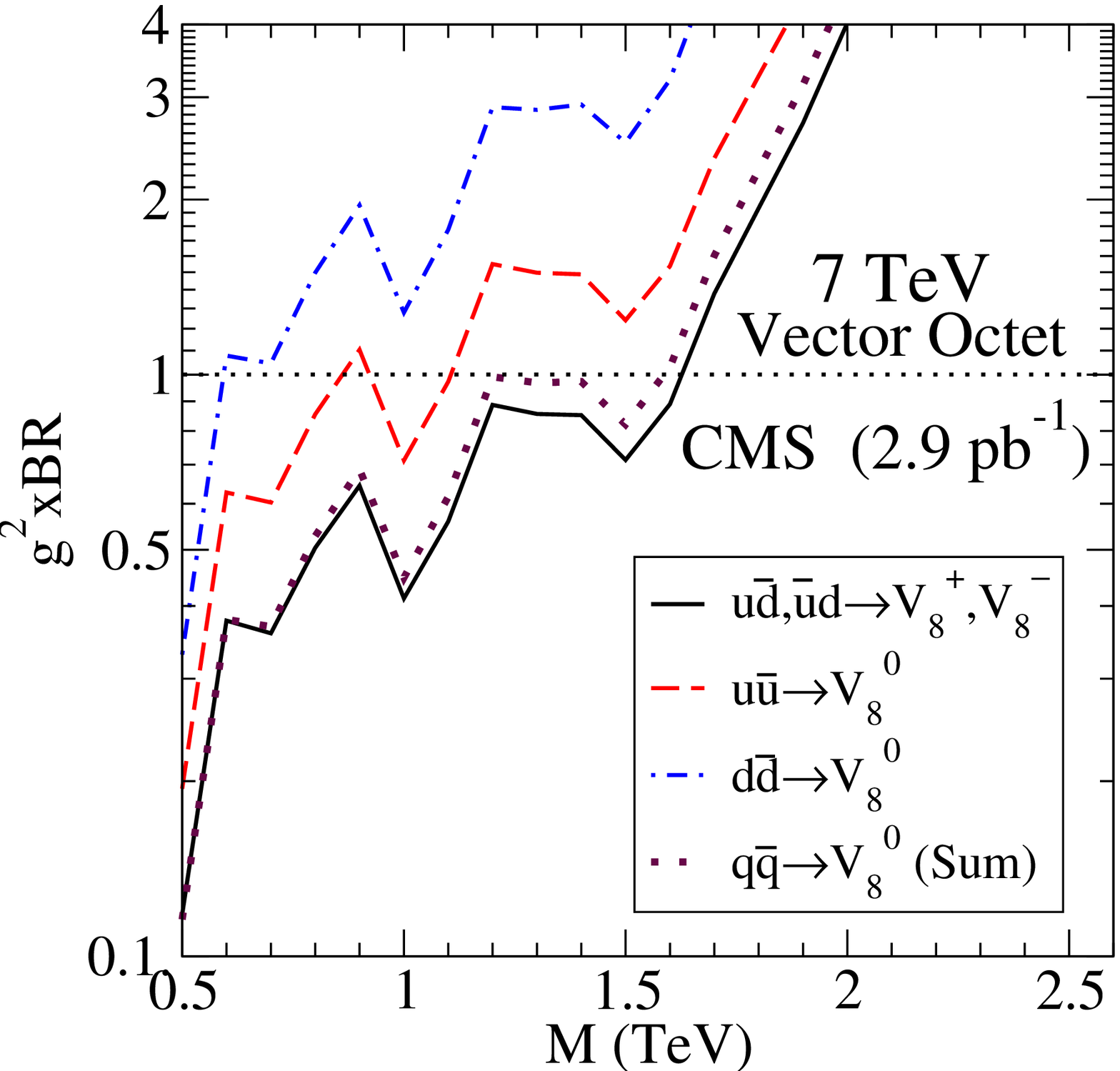}}
\caption{Bounds on the products of SM parton couplings to the resonances and dijet branching ratios 
(vertical axis) versus the resonant mass 
for (a) color-sextet vector diquarks, (b) color-sextet (or color-triplet) excited quarks, 
(c) color-octet scalar, and (d) color-octet vectors from (a,b,c,d) CMS and (b) ATLAS.}
\label{bnds.FIG}
\end{figure}

All of the colored resonances presented in the previous sections
can contribute to the dijet signal; hence, the ATLAS and CMS dijet cross section bounds can be used to place limits on
the mass and couplings of these new particles.
We consider the current bounds on those colored resonances as summarized in Table \ref{qnum.tab2}.
In presenting our results for the current bound, we once again parameterize the signal rates by an overall coupling to the initial state partons and a branching fraction to decay to the final state dijets.
The limits on the product of the two constants of new colored resonances as a function of the resonant mass
are shown in Fig.~\ref{bnds.FIG}.  
The color-sextet vector diquark and color-octet scalar bounds are based on the leading order QCD calculations presented here with K-factors from QCD corrections, while all other bounds are based solely on the leading order calculations.
The regions above the corresponding curves are excluded, thus providing meaningful upper bounds for the
couplings and lower bounds for the resonant masses.
The zigzag shapes of the curves are due to the non-smooth experimental bounds for different masses as in
Fig.~\ref{CMS.FIG}.

Figure \ref{bnds.FIG}(a)\footnote{We have extended the $E^{\mu}_6$ bound beyond the CMS data point at $2.6$ TeV, assuming there has been no event observed.} 
shows the CMS bounds on the sextet vector diquarks with a NLO K-factor of 1.2 included \cite{Han:2009ya}.  The bounds on the scalar $U_6,E_6$ couplings are the same as those on vector $U^{\mu}_6,E^{\mu}_6$ and the bounds on the $D_6$ couplings are twice weaker than the bound on $D^{\mu}_6$. Furthermore, taking into account the different K-factors of 1.3 for triplets and 1.2 for sextets , the bound on the antisextet vectors are 1.8 times stronger than the bounds on the triplet vectors and the bounds on $D^{\mu}_6$ are 3.7 times stronger than $D_3$.

Figure \ref{bnds.FIG}(b) shows the bounds on the excited quarks.
Results based on the ATLAS data (solid curves) and CMS data (dashed curves) are comparable.
Following the convention in Ref.~\cite{Baur:1987ga}, we have set $\Lambda =2M$.
As noted earlier, the results for a color-triplet and sextet are the same. The bounds obtained here are stronger
than those for the diquarks above.

Figure \ref{bnds.FIG}(c) shows the bounds on octet scalar couplings including NLL K-factors running from $2.4$ at resonance mass $0.5$~TeV and $3.5$ at $2.5$~TeV~\cite{Idilbi:2009cc}.  Even with the K-factor, the weakest bound of all studied are the $gg$ initial processes.
This is due to the sharp fall of $gg$ luminosity at higher masses.  
The bounds on the coupling constants of the $T_8$ are a factor of two weaker than those of the $S_8$.

Although not as large as $uu,\ dd$ initial states, the $q\bar q$ annihilation provides reasonable sensitivity
to the colored resonances. In comparison with the Tevatron as a $p\bar p$ collider,
the LHC is somewhat in a disadvantageous situation with respect to the valence quark dominance.
Nevertheless, the LHC results currently have slightly extended the Tevatron bounds
on axigluons and universal colorons \cite{Collaboration:2010jd}.
We also obtain significant bounds for the color-octet resonances
based on the CMS data as seen in Fig.~\ref{bnds.FIG}(d).
However, due to the much larger data sample,  the Tevatron dijet bounds for color-singlet vectors
($Z',\ W'$ etc.) are much more stringent than those from the LHC.

Assuming a coupling constant and branching ratio of unity as indicated by the horizontal dotted lines
in Fig.~\ref{bnds.FIG}, the current mass lower bounds on the colored resonant states are summarized as 
\begin{equation*}
\begin{array}{rlrlrl}
E_6^{\mu}~~~&  2.7~\rm{TeV~(CMS)}~& \hskip -0.3in
E_6 ~~~&  2.7~\rm{TeV~}~~~ \\%
D_6^{\mu}~~~& 2.3~\rm{TeV~(CMS)} &
D_6 ~~~&  1.9~\rm{TeV~}~~~ \\
U_6^{\mu}~~~ & 1.8~\rm{TeV~(CMS)} & ~~~~~ U_6 ~~~&  1.8~\rm{TeV~}~~~\\ \\
E^{\mu}_3~~~ & 2.5~\rm{TeV~(CMS)} & ~~~~~ U^{\mu}_3 ~~~ &0.8,1.0-1.2,1.4-1.6 \rm{TeV}\\
D_3^{\mu} ~~~&  1.9~\rm{TeV~(CMS)}~~~ & D_3 ~~~&  0.8,0.9-1.2,1.3-1.7~\rm{TeV~} \\ \\
u_{6}^*~~~ & 1.7~\rm{TeV~(CMS)},~1.6~\rm{TeV~(ATLAS)} & d_{6}^*~~~ & 1.1~\rm{TeV, ~1.2~\rm{TeV }} \\ \\
V_8^{\pm}~~~ & 1.6~\rm{TeV~(CMS)}  & V_8^{0} ~~~ & 1.6~\rm{TeV~} \\ \\
S_8~~ & 1.2~\rm{TeV~(CMS)} & T_8~~~ & 0.9~\rm{TeV~},
\end{array}
\end{equation*}
where the mass bounds have been rounded to the nearest tenth of a TeV. 
All of the bounds obtained here are beyond the existing Tevatron analyses. 

It should be noted that there are small uncertainties associated with the results above.  For instance, the bounds presented above have been obtained by utilizing the narrow width approximation.  Also, the detector acceptances are somewhat dependent on the spin of the resonance.

\section{Conclusion}
\label{sec:conc}

Experiments at the LHC have opened up the energy frontier for TeV scale new physics searches.
Motivated by the recent ATLAS and CMS dijet analyses, we study the possible colored resonances
in a most general approach.
We classify the colored resonances based on group theory decomposition of QCD $SU(3)_{C}$
interaction as well as other quantum numbers, as listed in Table  \ref{qnum.tab}.
These resonances may carry exotic SM quantum numbers, but all of them find their interesting roles
in certain theories beyond the SM.

We then construct their effective couplings to light partons. Based on those features,
we name them and list them in Table  \ref{qnum.tab2}. 
We calculate their resonant production cross section at the LHC.
The production rates may be as large as 
400 pb (1000 pb) at the c.m.~energy of  7  (14) TeV for a mass of 1 TeV, 
leading to the largest production rates for new physics at the TeV scale, and simplest event
topology with dijet final states.  
Our approach is quite general and applicable to other possible signals of resonant particles other than dijets at the LHC.

We applied the new ATLAS/CMS dijet data to have put bounds
on various possible colored resonant states. We obtained the lower bounds on their
masses ranging from 0.9 to 2.7~TeV,  if their couplings are of the order of unity. 
The results  obtained here are beyond the existing Tevatron analyses. 
In an optimal situation, if a signal above the SM backgrounds is established in the near future, 
it is then the exciting time to determine the nature of the resonance particle and to untangle the 
new underlying theory as commented in the text and in Table \ref{qnum.tab2}.
With the anticipated increase of integrated luminosity and c.m.~energy,
experiments at the LHC will undoubtedly take our understanding of particle physics to an
unprecedented level.

\section{Acknowledgement}
We would like to thank Georgios Choudalakis, Fabio Maltoni, Tim Tait, and Brian Yencho for discussions. I.L. is supported by a DOE Graduate Fellowship in High Energy Theory.  This work was supported in part by the US DOE under contract No.~DE-FG02-95ER40896.

\appendix
\section{Clebsch-Gordan Coefficients}
\label{color.app}
Here we exhibit the Clebsch-Gordan Coefficients
relations that are needed in color resonance calculation for general
$SU(N)$ groups. Typical initial state group structure includes
$N\otimes N $, $N\otimes \bar{N}$, $N\otimes A$ and $A\otimes A$,
whereas A denotes the adjoint representation of $SU(N)$.

We assume that the Clebsch-Gordan coefficients obey the following
orthogonality relationship:
\begin{eqnarray}
\Tr[K^a\bar{K}_b]=\delta^a_b,
\end{eqnarray}
where $K^a=\bar{K}^\dagger_a$.  The Clebsch-Gordan coefficients couple the different irreducible representations together in a gauge invariant way.  Hence, depending on the interaction, the indices $a$ take on different values.  For example, according to our conventions in Eqs.~(\ref{eq:qq}) and (\ref{eq:qstar}):
\begin{equation}
a=\left\{\begin{array}{r l} 1,...,\frac{N(N-1)}{2} & {\rm~couplings~to~the~antisymmetric~combination~of}
~N \otimes N~{\rm boson}\\
                                            1,...,\frac{N(N+1)}{2} & {\rm~couplings~to~the~symmetric~combination~of}~N \otimes N~{\rm boson}\\
                                            1,...,N^2-1 & {\rm~couplings~to~the~new~fermion~from}~N\otimes A
                                            \end{array}\right .
\end{equation}
Using this orthogonality condition,
the following identities are used for the appropriate couplings:
\begin{equation}
\Tr[K^a\bar{K}_a]=\left\{\begin{array}{r l} \frac{N(N-1)}{2} & {\rm~couplings~to~the~antisymmetric~combination~of}
~N \otimes N~{\rm boson}\\
                                            \frac{N(N+1)}{2} & {\rm~couplings~to~the~symmetric~combination~of}~N \otimes N~{\rm boson}\\
                                            N^2-1 & {\rm~couplings~to~the~new~fermion~from}~N\otimes A
                                            \end{array}\right .
\end{equation}

Specifically for $SU(3)$ couplings, the color antisymmetric (symmetric) coupling of two quarks is the triplet-(sextet) scalar or vector diquark.  Then we have for the diquark interactions in Eq.~(\ref{eq:qq}) and excited quark interactions in Eq.~(\ref{eq:qstar})
\begin{equation}
\Tr[K^a\bar{K}_a]=\left\{\begin{array}{r l} 3 & \rm{~couplings~to~the~triplet~diquark}\\
                                            6 & \rm{~couplings~to~the~sextet~diquark}\\
                                            8 & \rm{~couplings~to~the~triplet~and~sextet~excited~quarks}
                                            \end{array}\right .
\end{equation}

The complete symmetric invariant symbol of $SU(3)$ algebra satisfy
\begin{equation}
d^{ABC}d^{ABC}=2 C_F (N^2-4)
\end{equation}
where as $C_F= (N^2-1)/ {2N}$ is
the eigenvalue of the quadratic $SU(N)$ Casimir operator acting on the fundamental representation.
For $SU(3)$, $C_F= 4/ 3$.

\section{Feynman Rules}
\label{app:Feynman}

Here we give the explicit Feynman rules for the interacting vertices constructed in the text,
as in
Sec.~\ref{sec:inter}.

\begin{figure}[tb]
\centering
\includegraphics[clip,width=0.8\textwidth]{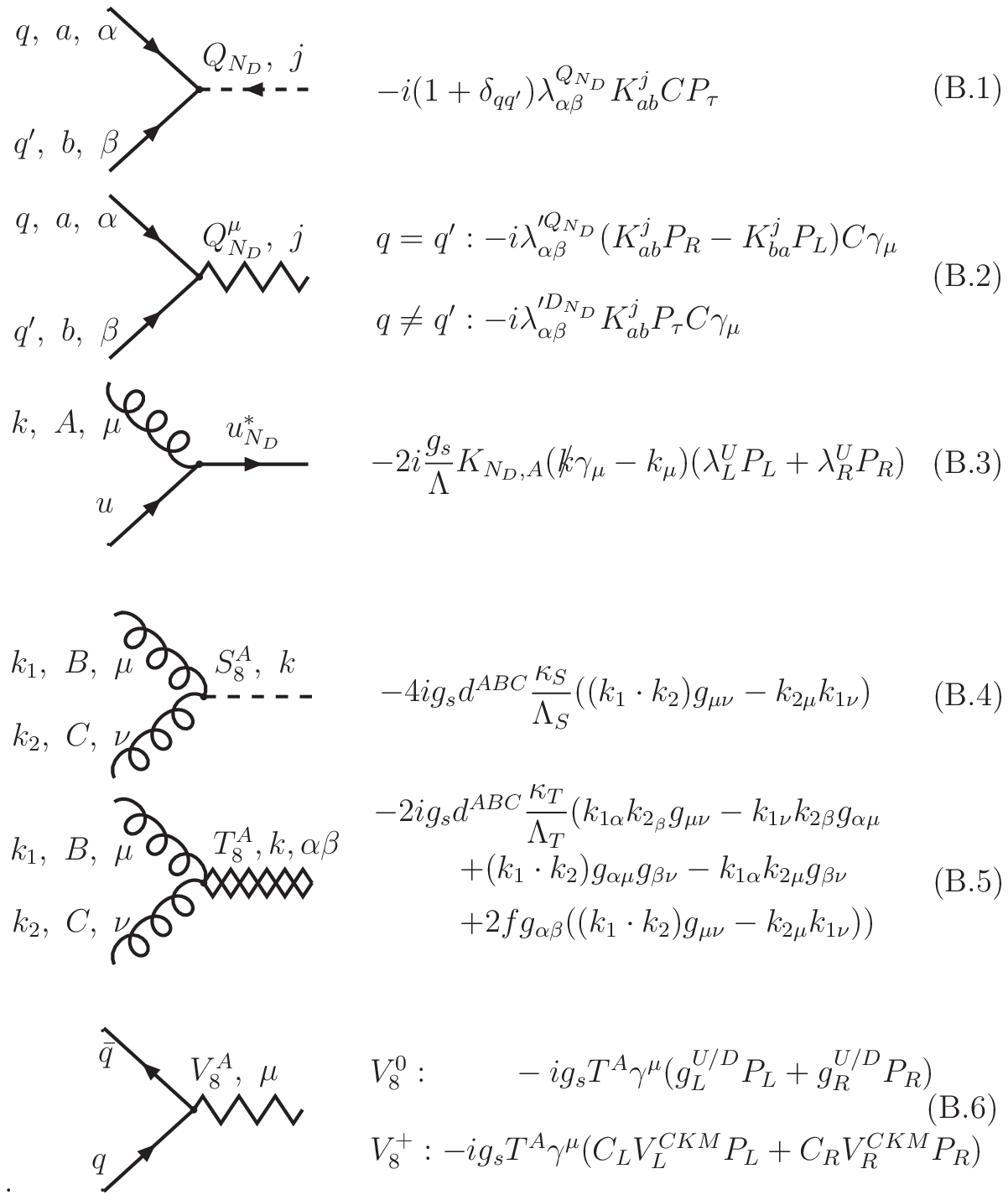}
\caption{Feynman rules for the vertices of resonant particle couplings to quarks and gluons.
All the momenta are incoming and $\tau=L,R$.}
\label{fig:Feyn}
\end{figure}

The diquark Feynman rules are in Eq.~(B.1) and (B.2): $Q_{N_D}$ may be $E_{N_D},D_{N_D},$ or $U_{N_D}$, depending on the initial state.  The labels $q,q'=u,d$ indicate whether the initial state quarks are up-type or down-type and $\alpha,\beta$ are generation indices.  For the triplet diquark the Clebsch-Gordan coefficients are antisymmetric $K^j_{ab}=-K^j_{ba}$, and for the antisextet diquark they are symmetric $K^j_{ab}=K^j_{ba}$. $C$ is the charge conjugation matrix.

The excited quark Feynman rules in Eq.~(B.3): equally applicable for both
$u^*_{N_D}$ and $d^*_{N_D}$. 

The spin summation for a spin-2 tensor state ($T_{8}^{A}$) of mass $M$
obeys the relation  \cite{Han:1998sg}:
\begin{eqnarray}
\Sigma\epsilon_{\mu\nu}(k)\epsilon^\ast_{\rho\sigma}(k) &=&
B_{\mu\nu,\rho\sigma}(k) = \left(g_{\mu\rho}-{k_\mu k_\rho\over
M^2}\right) \left(g_{\nu\sigma}-{k_\nu
k_\sigma\over M^2}\right)\nonumber \\
&+& \left(g_{\mu\sigma}-{k_\mu k_\sigma\over M^{2}}\right) 
\left(g_{\nu\rho}-{k_\nu k_\rho\over M^{2}}\right) - {2\over3}\left(g_{\mu\nu}-{k_\mu
k_\nu\over M^2}\right) \left(g_{\rho\sigma}-{k_\rho k_\sigma\over M^{2}}\right) . 
\nonumber
\end{eqnarray}

\end{document}